\def\year{2022}\relax
\documentclass[letterpaper]{article} 
\usepackage{aaai22}  
\usepackage{times}  
\usepackage{helvet}  
\usepackage{courier}  
\usepackage[hyphens]{url}  
\usepackage{graphicx} 
\usepackage{natbib}  
\usepackage{caption} 
\DeclareCaptionStyle{ruled}{labelfont=normalfont,labelsep=colon,strut=off} 
\usepackage{algorithm}
\usepackage{algorithmic}

\usepackage{microtype}
\usepackage{graphicx}
\usepackage{float}
\usepackage{caption}
\usepackage{todonotes}
\usepackage{dsfont}
\usepackage{wrapfig}
\usepackage{amsthm}
\usepackage{booktabs} 
\usepackage{xcolor}
\usepackage{newfloat}
\usepackage{listings}
\usepackage{appendix}

\usepackage{amsmath}
\usepackage{amsfonts}
\usepackage{subcaption}
\newtheorem{theorem}{Theorem}

\newcommand{\norm}[1]{\left\lvert#1\right\rvert}

\DeclareMathOperator*{\argmin}{arg\!\,min}

\newcommand{\rt}{$O(\frac{1}{\sqrt{T}})$\,}

\newcommand{\pyear}[1]{\citeauthor{#1} (\citeyear{#1})}

\floatstyle{ruled}
\newfloat{listing}{tb}{lst}{}
\floatname{listing}{Listing}
\title{Equilibrium Finding in Normal-Form Games Via Greedy Regret Minimization}
\iftrue
\author{
    Hugh Zhang\textsuperscript{\rm {1}\footnote{Work done primarily while the author was at Facebook AI Research.}},
    Adam Lerer\textsuperscript{\rm {2}},
    Noam Brown\textsuperscript{\rm {2}}
}
\affiliations{
    \textsuperscript{\rm 1}Harvard University, \textsuperscript{\rm 2}Facebook AI Research \\
    hughzhang@fas.harvard.edu, alerer@fb.com, noambrown@fb.com
}
\fi

\begin{document}

\maketitle

\begin{abstract}
We extend the classic regret minimization framework for approximating equilibria in normal-form games by greedily weighing iterates based on regrets observed at runtime. Theoretically, our method retains all previous convergence rate guarantees. Empirically, experiments on large randomly generated games and normal-form subgames of the AI benchmark Diplomacy show that greedy weights outperforms previous methods whenever sampling is used, sometimes by several orders of magnitude.
\end{abstract}

\section{Introduction}
\looseness=-1
Constructing algorithms that efficiently converge to equilibria is one of the central goals of computational game theory. In recent years, regret minimization techniques for approximating equilibria via self play have led to a number of major successes in games like poker~\citep{bowling2015heads,moravvcik2017deepstack,brown2018superhuman,brown2019superhuman}, Avalon \citep{serrino2019finding}, and Diplomacy \citep{gray2021human}. 
Regret minimization is now the state-of-the-art approach for computing equilibria in large games, especially those with a large number of actions or in settings where queries to the payoff matrix are too expensive to compute an exact solution via linear programming.

Classical usage of regret minimization algorithms for learning equilibria in games has typically weighted each iteration equally.
In this paper, we demonstrate empirically faster convergence, while still guaranteeing the same worst-case regret bound, by greedily weighing each new iteration to minimize the regret minimizer's potential function: a measure the current distance to the set of desired equilibria. Recent work such as CFR+~\cite{tammelin2014solving} and Linear CFR~\cite{brown2019solving} have also shown that faster performance is achievable by weighing iterates non-uniformly. However, in all previous algorithms, iterates were weighed according to a fixed, pre-determined schedule. In contrast, we introduce \textbf{greedy weights}, the first equilibrium-finding regret minimization algorithm where iterates are weighed \emph{dynamically} using information available at \emph{runtime}.
\begin{figure}[t!]
  \includegraphics[width=\linewidth]{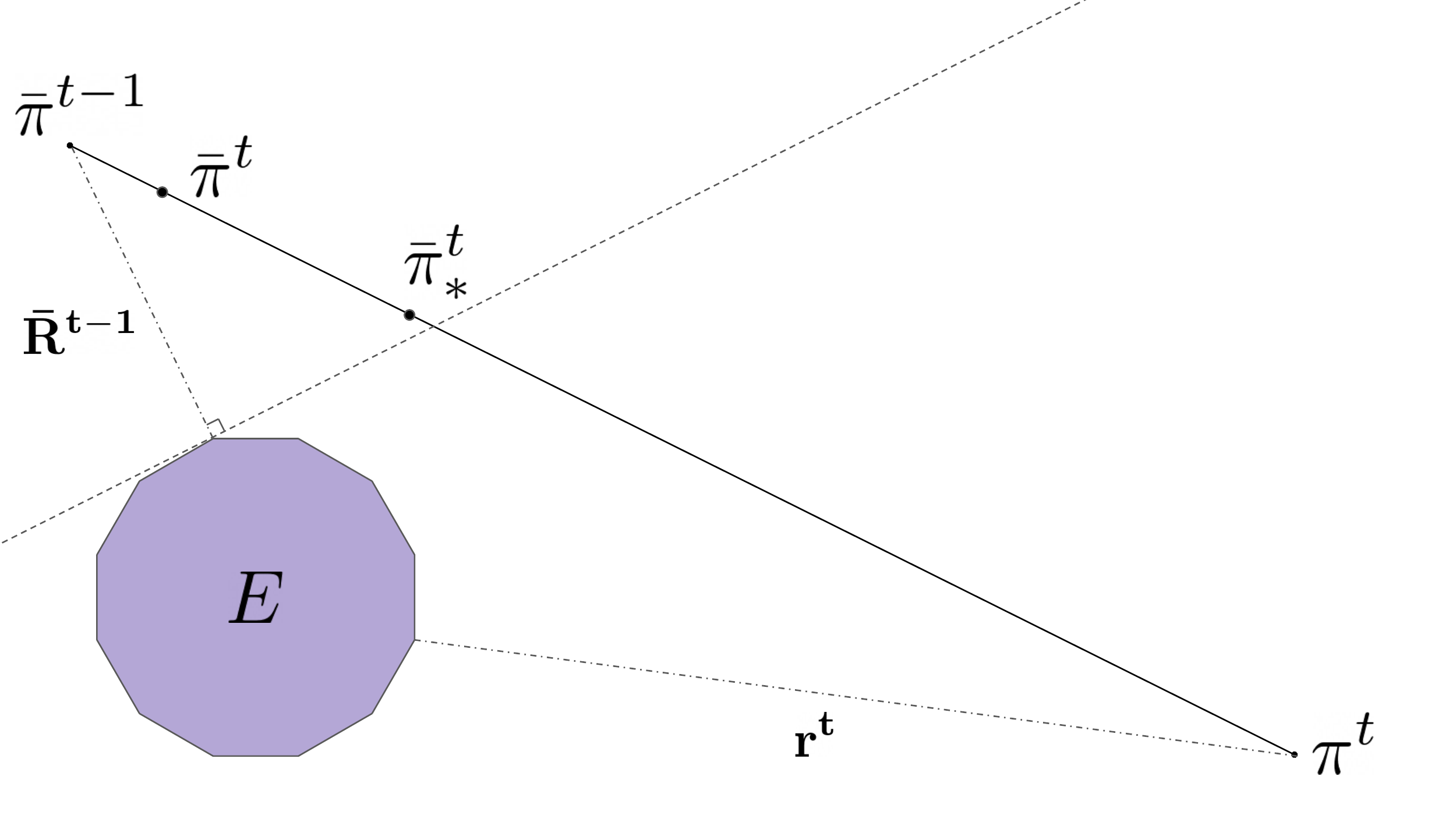}
  \caption{
  Vanilla regret minimization methods always move the rolling weighted average policy profile discovered so far $\bar{\pi}^{t-1}$ (with average regret represented by $\bf \bar{R}^{t-1}$) a fixed step towards the latest iteration of the procedure $\pi^t$ (with instantaneous regret represented by $\bf r^t$) to the point represented by $\bar{\pi}^t$. In contrast, greedy weights will choose the point $\bar{\pi}^t_*$ that minimizes the distance to the set of equilibria we wish to approach (represented by $E$), often resulting in accelerated convergence.
  }
  \label{fig:intuitive-graphic}
\end{figure}
We benchmark greedy weights against past techniques for computing minimax equilibria (in two-player zero-sum games), coarse-correlated equilibria, and correlated equilibria in large, randomly generated normal-form games. Additionally, we conduct experiments in subgames of Diplomacy, which has a long history as an important benchmark for AI research~\citep{kraus1988diplomat,kraus1994negotiation,kraus1995designing,johansson2005tactical,ferreira2015dipblue} and has been a particularly active domain for research in recent years~\citep{paquette2019no,anthony2020learning,gray2021human}.
We find that greedy weights significantly improves the rate of convergence whenever sampling is used, in some cases by several orders of magnitude. Finally, we find that equilibria discovered by greedy weights in general-sum games typically have higher overall social welfare than those found by prior methods.
\section{Notation and Background}
\label{sec:background}
In a normal-form (also called strategic-form) game, each of $\mathcal{P}$ players simultaneously chooses their actions without observing the other players' choices. Each player then receives a reward determined by a function of all players' actions. All games can be written as normal-form games, though some may additionally admit more compact representations. Let $\Delta$ represent the difference between the highest and lowest possible payoff of any player in the game. Let $A_i$ denote the set of actions for player~$i$ and $A$ the set of joint actions for all players.
We denote the set of joint actions for all players \emph{except} $i$ as  $A_{-i}$. Let $\Sigma_i$ represent the set of probability distributions over actions in $A_i$ (i.e., the set of mixed policies, also known as strategies). $\Sigma$ is the set of joint policies across all players, and $\Sigma_{-i}$ is the set of joint policies for all players other than $i$. $\pi_i \in \Sigma_i$ denotes player~$i$'s policy, which is a probability distribution over actions in $A_i$. The probability of action~$a_i$ in $\pi_i$ is denoted $\pi_i(a_i)$. Similarly, $\pi_{-i}$ and $\pi$ denote the policies for all players other than $i$ and for all players, respectively.
The payoff player $i$ receives when all players play joint action~$a\in A$ is denoted $u_i(a) = u_i(a_i, a_{-i})$. Analogously, the expected payoff to player $i$ when all players play policy profile $\pi$ is denoted $u_i(
\pi) = u_i(\pi_i, \pi_{-i})$.

\subsection{Equilibria in Games}
Perhaps the most well-known equilibrium concept for games is the Nash equilibrium (NE)~\cite{nash1951non}. A NE is a tuple of policies (one for each player) in which no player can improve by deviating to a different policy. Formally, a policy profile $\pi$ is a NE if it satisfies:
\begin{equation}
\max_{i \in \mathcal{P}} \max_{\pi^\prime \in \Sigma_i} \;u_i(\pi^\prime, \pi_{-i}) - u_i(\pi_i, \pi_{-i}) \le 0
\end{equation}
Well-known results in complexity theory have suggested that discovering (or even approximating) a NE in general games is computationally hard \citep{chen2009settling, daskalakis2009complexity, rubinstein2019hardness}. As a result, researchers often also consider the correlated equilibrium (CE)~\citep{aumann1974subjectivity}, an alternative solution concept which is efficiently computable in all normal-form games. Whereas a NE is a probability distribution over actions $A_i$ for each player~$i$, a CE is a probability distribution $p$ over the set of joint actions~$A$ that satisfies certain incentive constraints. In order for a probability distribution over joint actions to be a CE, it must be the case that if a mediator were to sample a joint action~$a$ from that distribution and privately share with each player~$i$ their action~$a_i$ that is part of the joint action, then no player could gain by deviating from that action. Formally, a CE is a probability distribution over joint actions in $A$ satisfying

\[
 \max_{i \in \mathcal{P}} \max_{\phi: A_i \rightarrow A_i} \sum_{a \in A} p(a) (u_i(\phi(a_i), a_{-i}) - u_i(a_i, a_{-i})) \le 0
\]

Finally, a \emph{coarse}-correlated equilibrium (CCE)~\citep{hannan1957approximation} is also a probability distribution over joint actions but with a weaker incentive constraint than the correlated equilibrium. In order for a probability distribution over joint actions to be a CCE, it must be the case that if a mediator were to sample a joint action~$a$ from that distribution and each player~$i$ is forced to play their action~$a_i$ that is part of the joint action, then no player could gain by refusing to receive an action from the mediator and choosing an action on their own instead. Formally, a CCE satisfies:
\begin{equation}
\max_{i \in \mathcal{P}} \max_{a'_i \in A_i} \sum_{a \in A} p(a) (u_i(a'_i, a_{-i}) - u_i(a_i, a_{-i})) \le 0
\end{equation}

We can define the $\epsilon$-versions of all of the above equilibria by replacing the $0$ on the right hand side of the equations with an $\epsilon$.
In two-player zero-sum games, NE, CE, and CCE can be shown to be payoff equivalent to one another via the minimax theorem ~\cite{neumann1928theorie}.

\subsection{Regret Minimization}
There exist several polynomial-time algorithms for computing CEs and CCEs. This paper focuses on the leading approach for large games: regret minimization algorithms. In addition to their theoretical guarantees of convergence to equilibria, regret minimization algorithms have been behind recent empirical successes in large-scale game benchmarks such as many forms of poker (including non-two-player poker)~\citep{bowling2015heads,moravvcik2017deepstack,brown2018superhuman,brown2019superhuman}, Avalon \citep{serrino2019finding}, and Diplomacy \citep{gray2021human}.

For any sequence of policies $\pi^1 \dots \pi^T$ in a game $G$, player $i$'s weighted \textbf{external regret} for not having played action $a'_i \in A_i$ is 
\begin{align*}
R_i^{E,T}&(a'_i) =
 \sum_{t=1}^T w_t\big(u_i(a'_i, \pi^t_{-i}) - u_i(\pi^t)\big)
\end{align*}
We can thus define the overall average external regret for player~$i$ as
\begin{equation}
\bar{R}_i^{E,T} = \max\limits_{a'_i \in A_i} \frac{R_i^{E,T}(a'_i)}{\sum_{t=1}^T w_t} \end{equation}


Analogously, we can define player $i$'s weighted \textbf{internal regret} for not swapping to action $a'_i$ every time she actually played action $a_i^A$ as 
\begin{align*}
R_i^{I,T}(a^A_i, a'_i) =\sum_{t=1}^T \mathds{1}[a^t_i = a^A_i]w_t(u_i(a'_i, a^t_{-i}) - u_i(a^t))
\end{align*}
and her overall average internal regret as
\[
\bar{R}_i^{I,T} = \max\limits_{a'_i, a^A_i \in A_i} \frac{R_i^{I,T}(a^A_i, a'_i)}{\sum_{t=1}^T w_t}
\]
We denote the vector (for all players) of average external regrets by ${\bf \bar{R}}^{E,T}$ and average internal regret by ${\bf \bar{R}}^{I,T}$. Additionally, whenever we use the notation $R_+$, we refer to the positive regrets defined by $\max\left(0, R\right)$ for whatever regret $R$ represents. For example, $R^{E,t}_{i,+}(a_i) = \max\left(0, R_i^{E,t}(a_i)\right)$.

Celebrated past results have shown that for any strategic-form game, minimizing average external regret for all players leads to a CCE while minimizing average internal regret for all players leads to a CE \citep{cesa2006prediction}.
Many well-known methods exist for minimizing both internal and external regret. For our experiments in external regret minimization, we use Blackwell's regret minimization \citep{blackwell1956analog}, which asks every player to select their next action proportional to how much they regret having not selected that action in the past.
Formally, each player selects action $a_i \in A_i$ at timestep $t+1$ with probability
\[
  \Pr(a^{t+1}_i = a_i) = \frac{R^{E,t}_{i, +}(a_i)}{\sum_{a'_i \in A_i} R^{E,t}_{i,+}(a'_i)}
\]
except in the case where all regrets are nonpositive upon which $a^{t+1}_i$ is chosen uniformly at random from $A_i$.

For internal regret minimization, we primarily use an extension of Blackwell's regret minimization given by \pyear{hart2000simple} also known as \textbf{regret matching}. Regret matching also selects its policy with probability ``matching'' its past regrets of not switching to that action in the past, but it differs in that it additionally uses a fixed inertia parameter $\alpha$ and thus always retains a positive probability of staying in place, with probability approaching $1$ as the overall regrets vanish\footnote{Note that our formulation is slightly different from the formulation given in their original paper in that it allows us to use a very low inertia constant ($\alpha = 10^{-10}$) and thus prevents the procedure from repeating actions until the regrets become very low. We empirically observe that this dramatically accelerates convergence to CE in our experiments with large random games.}. Formally, we have
\[
  \Pr(a^{t+1}_i = a_i) = \left.
  \begin{cases}
    \frac{\alpha}{\alpha + \sum_{a'_i \in A_i} R^{I,t}_{i,+}(a^t_i, a'_i)}, & \text{if } a_i = a^t_i \\[10pt]
    \frac{R^{I,t}_{i,+}(a^t_i, a_i)}{\alpha + \sum_{a'_i \in A_i} R^{I,t}_{i,+}(a^t_i, a'_i)}, & \text{otherwise}
  \end{cases}
  \right\}
\]

Due to the substantial similarities between the two algorithms, we refer to both
as regret matching. Both forms of regret matching have been shown to be special cases of a general class of potential-based minimizers \citep{cesa2006prediction, hart2001general, cesa2001potential}. Specifically, many of their theoretical properties can be proved via careful examination of the potential function, defined as the sum of the squared positive regrets. Formally, the potential functions that regret matching minimizes for external and internal regret, respectively, are:
\[
\phi({\bf \bar{R}}_+^{E,T}) = \sum_{i \in P} \sum_{a_i \in A_i} \big(\bar{R}_{i,+}^{E,T}(a_i)\big)^2
\]
\[
\phi({\bf \bar{R}}_+^{I,T}) = \sum_{i \in P} \sum_{a^A_i, a^B_i \in A_i} \big(\bar{R}^{I,T}_{i,+}(a^A_i, a^B_i)\big)^2
\]

External regret matching guarantees that $\phi({\bf \bar{R}}_+^{E,T}) \leq \frac{\norm{P}\Delta^2 |A|}{T}$, which in turn guarantees that $\max\limits_{i \in P} \max\limits_{a_i \in A_i} \bar{R}_i^{E,T}(a_i) \leq \frac{\Delta \sqrt{\norm{P} \norm{A}}}{\sqrt{T}}$, with similar guarantees for internal regret. If all players' average regret is bounded by $\epsilon$, then the empirical distribution of play is an $O(\epsilon)$-equilibrium.
\section{Greedy Weights}
\label{sec:approach}
\label{sec:greedyweights}
Blackwell's original regret minimization procedure \citep{blackwell1956analog} and its various extensions (e.g. \pyear{hart2000simple}, \pyear{blum2007external}, \pyear{zinkevich2008regret}) and applications to online learning \citep{abernethy2011blackwell} typically assign equal weight to each iteration of the procedure. Recently, \pyear{brown2019solving} demonstrated that modifying the weight schedule of regret matching empirically resulted in faster convergence to equilibria while maintaining a similar worst-case convergence bound. However, this modified schedule was fixed and pre-determined before the start of the procedure. Our algorithm extends this direction of inquiry by greedily choosing the iteration weights to minimize a function of regret at runtime.

Figure~\ref{fig:intuitive-graphic} demonstrates why this might prove useful. We represent the set of equilibria we wish to approach as $E$. At iteration $t$, let $\bar{\pi}^{t-1}$ represent the weighted average policy thus far
and $\pi^t$ the policy played at iteration $t$ of the procedure. Their regrets are denoted by $\bf \bar{R}^{t-1}$ and $\bf r^t$, respectively. Vanilla regret minimization would give an overall weight of $\frac{1}{t}$ to iteration $t$, regardless of whether an alternative weighting would result in an average policy closer to an equilibrium.
In contrast, greedy weights would choose the relative weighting between $\bar{\pi}^{t-1}$ and $\pi^t$ that
minimizes the potential function measuring the distance of the resulting new average policy $\bar{\pi}^t_*$ to the set of equilibria $E$. 
Formally, greedy weights picks the weight of each iteration to greedily minimize $\phi_i^{E,T}$ or $\phi_i^{I,T}$, depending on whether external or internal regret is being minimized.

\begin{algorithm}[t]
   \caption{Greedy Weights}
\begin{algorithmic}
   \STATE {\bfseries Input:} total timesteps $T$, game $G$, regret minimizer $M$
   \STATE Initialize $a$ randomly
   and compute immediate regret ${\bf r}$
   \STATE Set $\bar{\pi} = a$, $w_{\text{sum}} = 1$, ${\bf R} = {\bf r}$
   \FOR{$t=1$ {\bfseries to} $T$}
      
       \STATE $\pi, {\bf r} \gets M(G, \phi, {\bf R})$
       \STATE $w \gets \argmin\limits_{w} \phi \left(({\bf R} + w {\bf r})/(w_{\text{sum}} + w)\right)$
       \STATE ${\bf R} \gets {\bf R} + w {\bf r}$
       \STATE $\bar{\pi} \gets (w_{\text{sum}} \bar{\pi} + w \pi) / (w_{\text{sum}} + w)$
       \STATE $w_{\text{sum}} \gets w_{\text{sum}} + w$
   \ENDFOR
   \STATE \textbf{return} $\bar{\pi}$
\end{algorithmic}
\label{mainalgorithm}
\end{algorithm}
Algorithm~\ref{mainalgorithm} provides a formal description for the greedy weights procedure. On each iteration the regret vectors for all players, denoted by ${\bf R}$, is used to determine the policy profile $\pi$ played in the regret minimization algorithm and the resulting instantaneous regret vectors for all players, denoted by ${\bf r}$ (which in typical regret minimization would be added to ${\bf R}$, leading to the new regret vectors). Next, the weight $w$ for ${\bf r}$ that minimizes the potential function is computed. Since the potential function is convex, this can be done either analytically or via a simple line search procedure. Both the update to the regret vectors and the update to the average policy profile is weighed by $w$, and the process repeats.
To reduce the risk of numerical instability and overflow, if $w > 1$, then rather than weigh the next iteration by $w$ we instead discount all previous iterations by $\frac{1}{w}$ and weigh the next iteration by $1$.

As with many greedy algorithms, greedy weights is not guaranteed to converge faster than vanilla regret matching.
In particular, we observe in two-player zero-sum games that setting a weight floor of $\frac{w_\text{sum}}{2t}$ is often useful for speeding up convergence. In all other settings, we did not observe a floor to be beneficial. In the Appendix, we 
describe ablations measuring the performance for different weight floors.

Computing the optimal weight is essentially a line search procedure. It can be approximated via binary search or computed exactly by checking $O(|\mathcal{P}||A|)$ points. This is especially useful in cases where ${\bf r}$ may be expensive to compute (e.g. when evaluated using a neural network value function), because this line search only requires a single evaluation of ${\bf r}$ and is then able to compute $\phi({\bf R} + w{\bf r})$ as a simple algebraic function of ${\bf R}, {\bf r}, w$ without any further queries to the reward function.

Greedy weights retains the convergence guarantees of previous regret minimization methods, as demonstrated by the following theorem.
\begin{theorem}
\label{mainproposition}
If the policies for each player and weights for each iteration are selected according to Algorithm~\ref{mainalgorithm} and run for $T$ iterations, the resulting average distribution of plays $\bar{\pi}^T$ is guaranteed to be an \rt-equilibrium.
\end{theorem}
The full proof for external and internal regret matching is provided in the Appendix.
\section{Experimental Results}
\label{sec:experiments}
\begin{figure}[b!]
  \includegraphics[width=\linewidth]{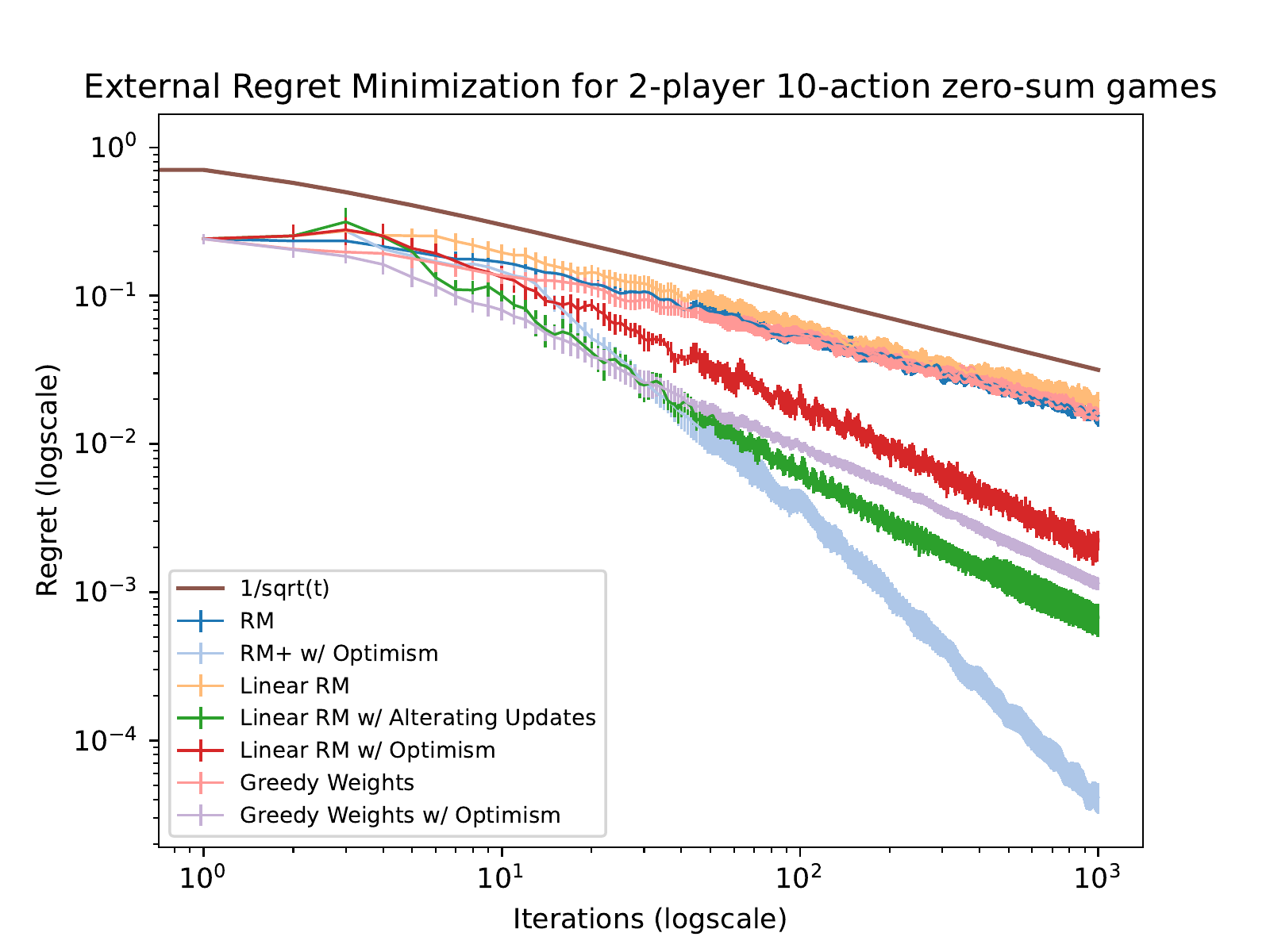}
  \caption{
  In the special case of two-player zero-sum games where mixed strategies are used at each iteration is applied, greedy weights (combined with optimism \citep{syrgkanis2015fast}) outperforms many but not all previous methods for minimizing external regret. However,
  this trick is not feasible for general computation of equilibria or in games with a large number actions where full queries to the payoff matrix are too expensive.
  }
  \label{fig:mixed-2pzs}
\end{figure}
We benchmark \textbf{greedy weights} against the state-of-the-art algorithms for regret minimization on randomly generated games and on subgames from the benchmark seven-player game Diplomacy.
In our experiments
we evaluate the following regret minimization methods:
\begin{figure*}[t!]
     \centering
     \begin{subfigure}[b]{0.49\textwidth}
         \centering
         \includegraphics[width=\textwidth]{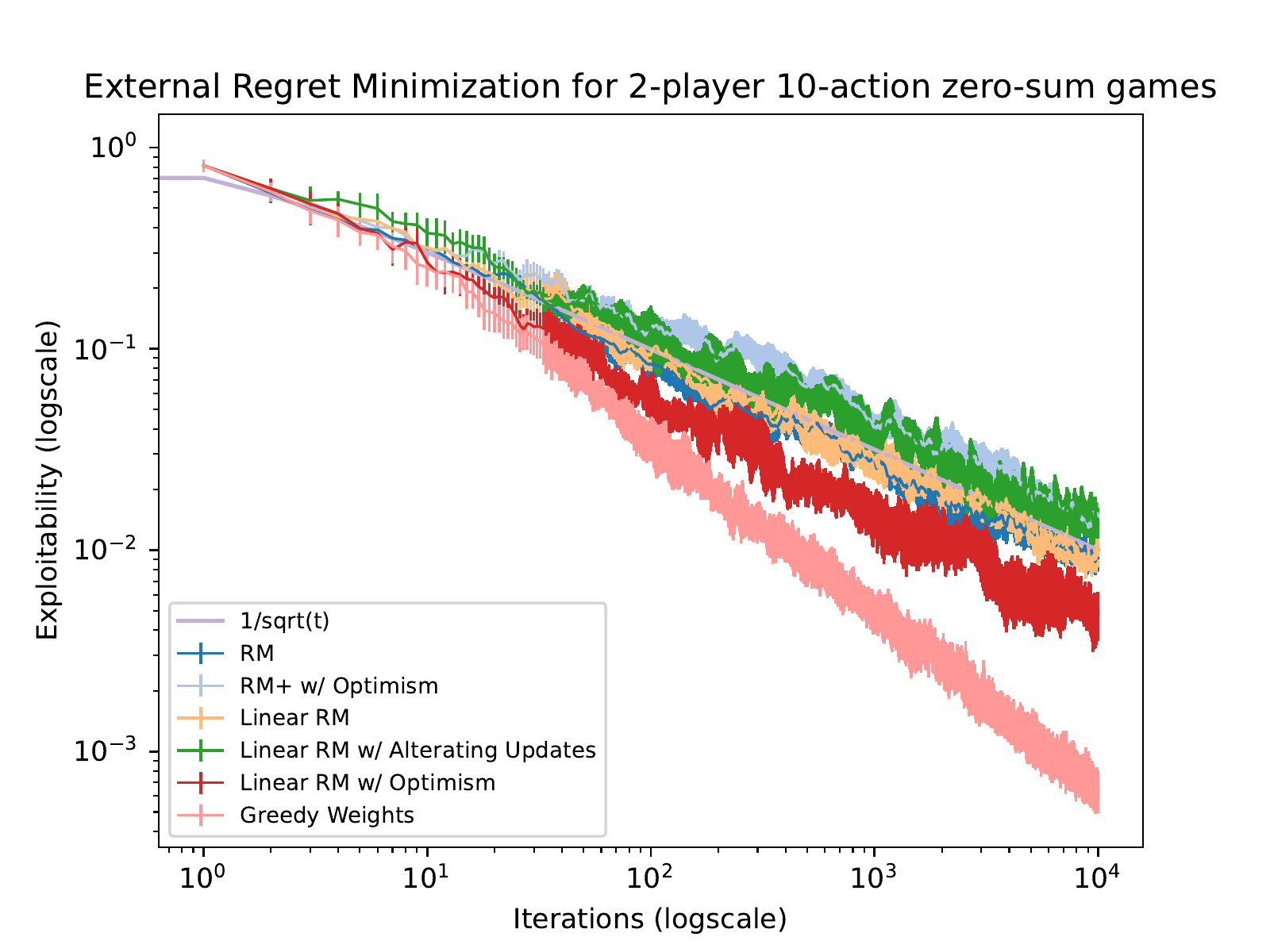}
         \caption{Exploitability v. Iteration}
         \label{fig:2pzs_regret_v_iteration}
     \end{subfigure}
     \hfill
     \begin{subfigure}[b]{0.49\textwidth}
         \centering
         \includegraphics[width=\textwidth]{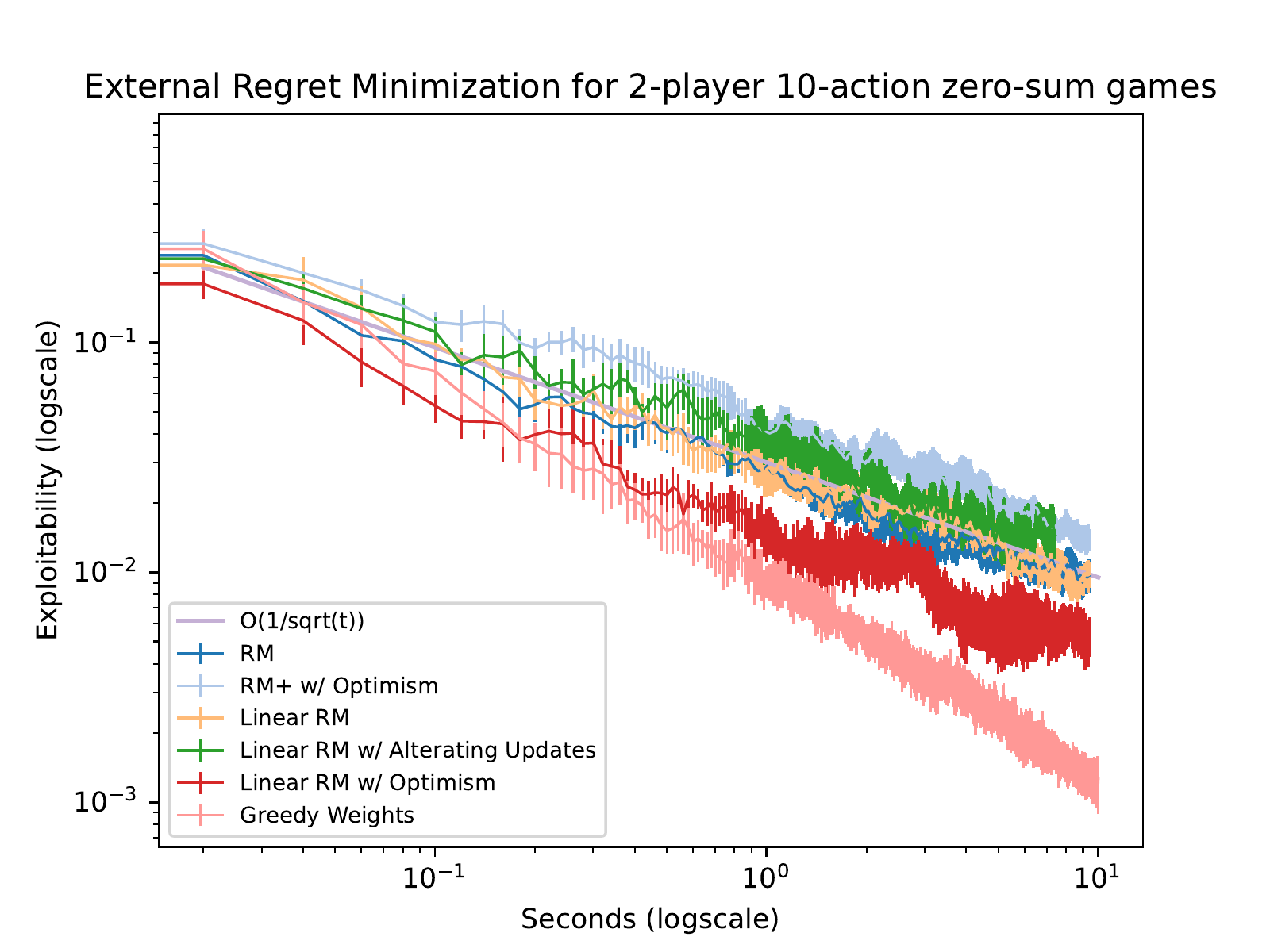}
         \caption{Exploitability v. Time}
         \label{fig:2pzs_regret_v_time}
     \end{subfigure}
     \hfill
        \caption{ We generate $10$ two-player zero-sum games with P1's matrix payoff entries selected uniformly at random from $[0, 1)$ and run the best of the described approaches for minimizing external regret in the sampled setting. Greedy weights outperforms all methods by almost an order of magnitude. Figure~\ref{fig:2pzs_regret_v_iteration} shows the results as a function of the number of iterations used, while Figure~\ref{fig:2pzs_regret_v_time} shows the same data as a function of time for fair comparison as iterations of greedy weights cost slightly more than previous methods. Exploitability is the distance to a Nash equilibrium. Both axes are logscale and error bars are shown at $95\%$ confidence.}
  \label{fig:2PZS}
\end{figure*}
\begin{enumerate}
    \item \textbf{Regret Matching (RM)} \cite{blackwell1956analog, hart2000simple}. Blackwell's was the the original (external) regret minimization procedure, where each player chooses their next policy proportionally to their positive regrets. \pyear{hart2000simple} extended their procedure to internal regret by adding a small inertia parameter that causes players to tend to stay in place as their regrets go to $0$.
    \item \textbf{RM+} \citep{tammelin2014solving}. RM+ makes two changes to RM. The first is that a distinction is made between the ``true'' regrets and the ``guiding'' regrets, where the guiding regrets are used to determine the policies on the next iteration. After every iteration, any negative guiding regrets are set to zero. The second is that in RM+ iteration $t$ is given weight $t$ when computing the final average policy (but not when computing the next iteration's policy). RM+ does well with mixed strategies, but does poorly with sampling~\citep{burch2017time}.
    \item \textbf{Linear RM} \citep{brown2019solving}. Linear RM is identical to vanilla RM except that iteration $t$ is given weight $t$ (both when computing the average policy and when computing the next iteration's policy).
\end{enumerate}
Additionally, prior work has discovered several additional modifications to regret minimization that significantly improves the convergence rate in two-player zero-sum games in practice. We additionally evaluate these methods.
\begin{enumerate}
    \item \textbf{Alternating Updates} \citep{tammelin2014solving}. For two-player zero-sum games, each player's guiding regrets are updated only once every other iteration. We generalize this procedure to $n$-player games by updating each player's regrets only once every $n$ iterations.
    \item \textbf{Optimism} \citep{syrgkanis2015fast, farina2021faster}. The guiding regrets are modified such that the latest iteration is temporarily counted twice. This boost is subtracted away from the guiding regrets immediately after the next iteration's strategy is determined and the equilibrium regrets remain unchanged.
\end{enumerate}
\subsection{Pure vs Mixed Policies at each Iteration}
\label{sec:mixedvspure}
In order to interpret the results in the following sections, it is necessary to discuss a technique often used in regret matching in the two-player zero-sum setting. When computing a Nash equilibrium in two-player zero-sum games, it is possible to simulate each player playing their \textit{mixed} policy at each iteration rather than sampling a single action to play. This modification leads to much faster convergence with certain regret matching schemes.
However, using mixed policies comes with some a major computational drawback in games with more than two players, as the computational cost of each iteration scales as $O(\norm{P}\norm{A}^{\norm{P}})$ for mixed-strategy RM as opposed to $O(\norm{A}\norm{P})$ for standard RM, making it intractable for games with more than 2-3 players, games with a large action space, or games where evaluating the payoff matrix is expensive. A game such as Diplomacy - an important AI challenge problem that has previously been tackled with RM \cite{gray2021human} - has all three of these properties, making mixed-strategy RM untenable.
We show that while some regret matching variants outperform greedy RM in the mixed-strategy two-player zero-sum setting (see Figure~\ref{fig:mixed-2pzs}), only greedy regret matching improves convergence in the general setting where mixed-strategy RM cannot be applied.

\subsection{Two-Player Zero-Sum Random Games}
We first evaluate greedy weights on the special case of two-player zero-sum games, where external regret minimization finds a Nash equilibrium. The results are shown in Figure~\ref{fig:2PZS}. Unlike the mixed case depicted in Figure~\ref{fig:mixed-2pzs}, no prior methods achieve a faster asymptotic convergence rate than \rt in the pure strategy setting. However, greedy RM empirically displays asymptotically faster convergence, and achieves regret over an order of magnitude lower after around $10^3$ iterations, even when accounting for the additional computational overhead of computing the optimal iteration weight. All experiments on random games (both zero-sum and general-sum) were done on a single CPU core.
\label{sec:generalgame}
\begin{figure*}
     \centering
     \begin{subfigure}[b]{0.49\textwidth}
         \centering
         \includegraphics[width=\textwidth]{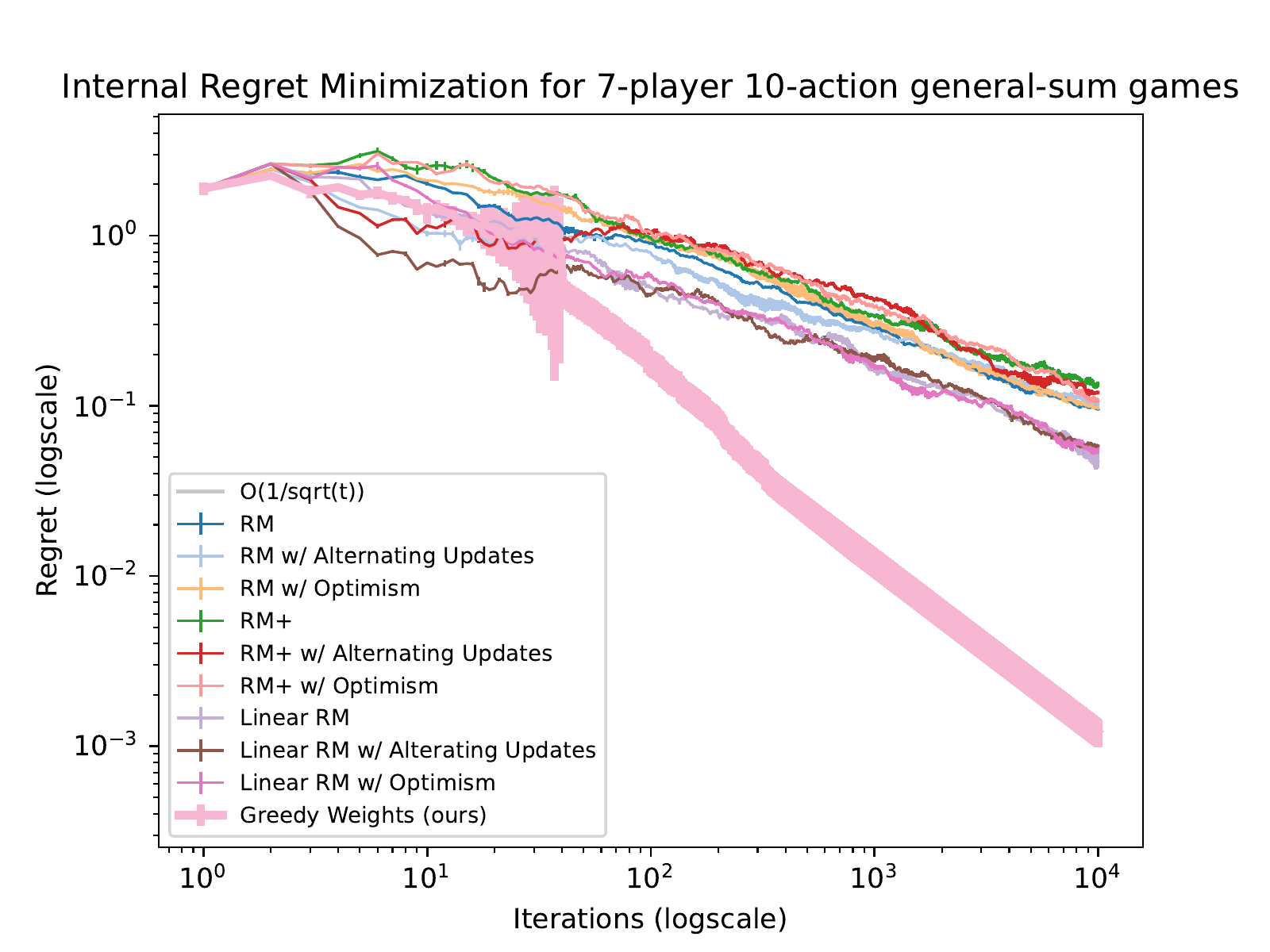}
         \caption{Regret v. Iteration}
         \label{fig:general_regret_v_iteration}
     \end{subfigure}
     \hfill
     \begin{subfigure}[b]{0.49\textwidth}
         \centering
         \includegraphics[width=\textwidth]{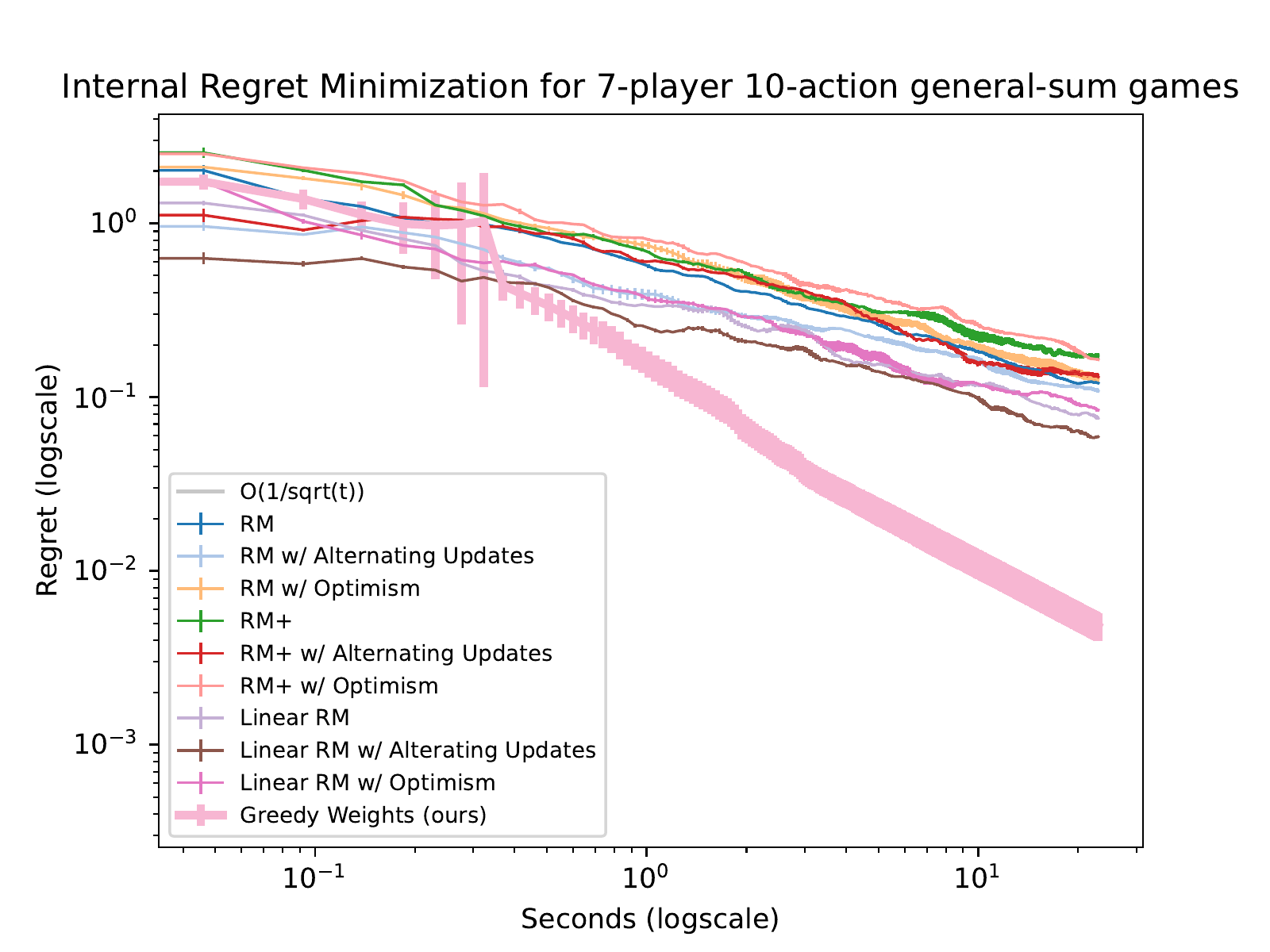}
         \caption{Regret v. Time}
         \label{fig:general_regret_v_time}
     \end{subfigure}
     \hfill
        \caption{We generate $10$ seven-player general-sum games with the matrix payoff entries selected uniformly at random from $[0, 1)$ and run all the described state-of-the-art approaches for minimizing internal regret in the sampled setting. All methods except our newly proposed greedy weights roughly converge at the worst case bound rate of \rt, while greedy weights converges several orders of magnitude faster. Figure~\ref{fig:general_regret_v_iteration} shows the results as a function of the number of iterations used, while Figure~\ref{fig:general_regret_v_time} shows it as a function of time for fair comparison as each iteration of greedy weights costs slightly more than previous methods for regret minimization. Note that both axes are logscale. Error bars are shown at $95\%$ confidence.}
  \label{fig:mainresult}
\end{figure*}
\subsection{General-Sum Games}
In Figure~\ref{fig:mainresult}, we compare the convergence of different RM weighting schemes for computing correlated equilibria in random games. We observe again that no prior methods achieve asymptotically faster convergence to correlated equilibria than \rt in general-sum games in the pure strategy setting. Our proposed greedy weighting scheme, however, dramatically improves convergence to correlated equilibria in large general-sum games. These results are robust to variance between games (error bars listed at $95\%$ confidence) and hold across a large spectrum of games with varying numbers of players and actions. For the sake of space, we have relegated most of these plots to the Appendix. Code to replicate the random normal-form game experiments can be found at https://github.com/hughbzhang/greedy-weights.

We also investigate whether the equilibria that greedy weights discovers in general-sum games increase total expected value (i.e., social welfare) compared to equilibria found using vanilla regret minimization. Intuitively, we might expect this to be the case since iterations where players receive higher overall rewards are also likely to be iterations with lower overall regret, and would thus be upweighted by the greedy weights algorithm.

We generated 100 random 7-player 10-action general-sum games with payoff entries randomly sampled between 0 and 1 and ran both the standard and greedy weights variants of regret matching for $1000$ iterations each. Indeed, greedy weights RM converges to equilibria with higher social welfare: it finds equilibrium with average welfare of $4.16 \pm 0.023$, while vanilla RM finds equilibria with average welfare $3.50 \pm 0.005$ (95\% confidence intervals).

Additionally, we evaluate greedy weights on all the normal-form games included in the popular game theory library OpenSpiel \cite{lanctot2019openspiel}. For space reasons, these plots have been relegated to the appendix.




\subsection{Results in Diplomacy}
\label{sec:diplomacy}
In addition to running experiments on randomly generated matrix games, we also benchmark greedy weights on subgames of the benchmark game of Diplomacy.

Diplomacy is a popular seven-player zero-sum board game that involves simultaneous moves and both cooperation and competition. Players decide whom to support and whom to betray in pursuit of majority control of the board. Diplomacy has a long history as a benchmark for AI research~\citep{kraus1988diplomat,kraus1994negotiation,kraus1995designing,johansson2005tactical,ferreira2015dipblue} and has been a particularly active domain for research in recent years~\citep{paquette2019no,anthony2020learning,gray2021human}. Since the game state is fully observable each turn, and players act simultaneously, each turn in Diplomacy can be viewed as a normal-form game if there is a defined state value function. Moreover, since players are able to communicate before acting, it is possible for players' actions to be correlated.

\begin{figure*}[ht!]
\centering
\begin{subfigure}[b]{.5\textwidth}
  \centering
 \includegraphics[width=\columnwidth]{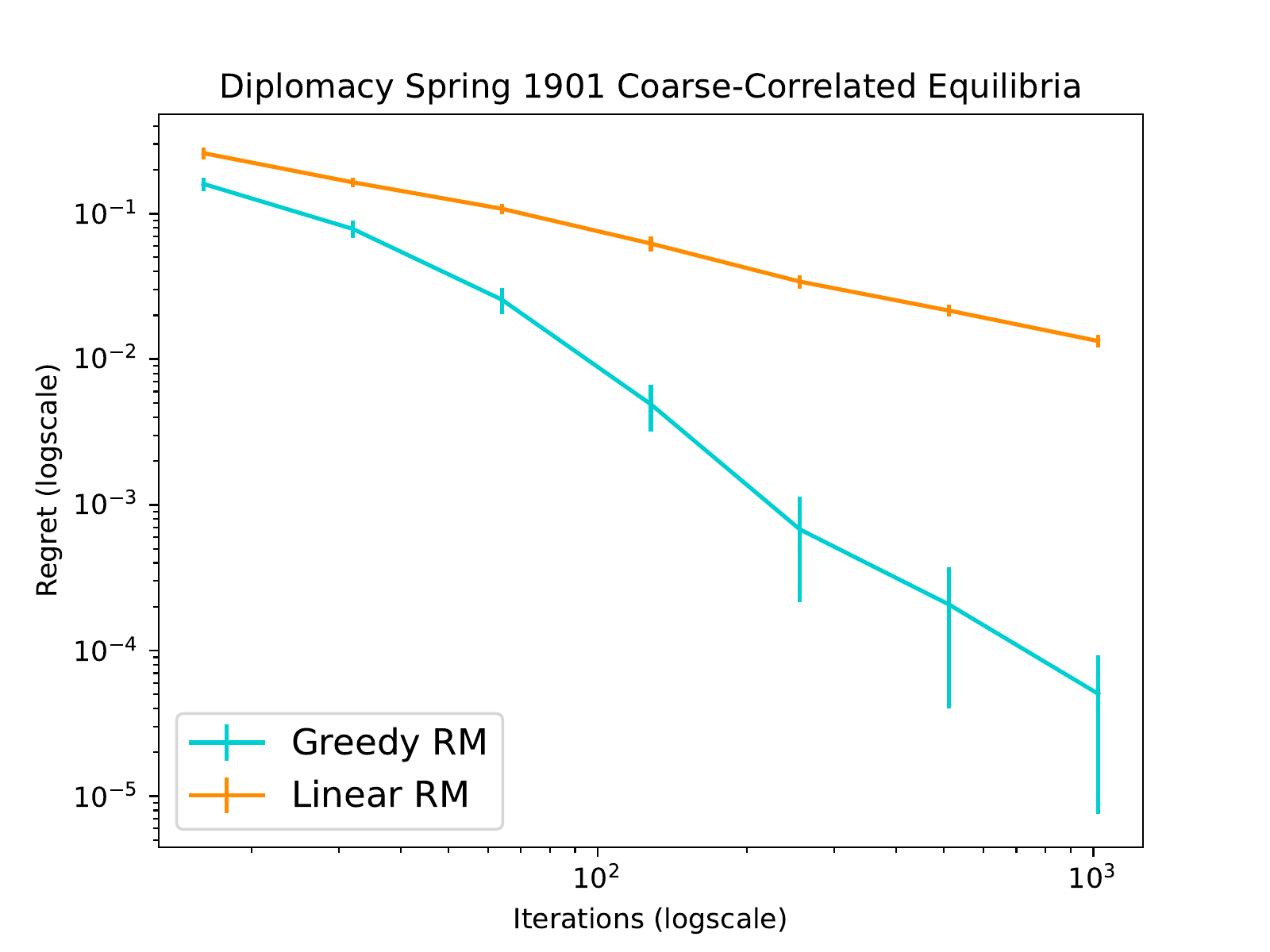}
\end{subfigure}%
\centering
\begin{subfigure}[b]{.5\textwidth}
  \centering
 \includegraphics[width=\columnwidth]{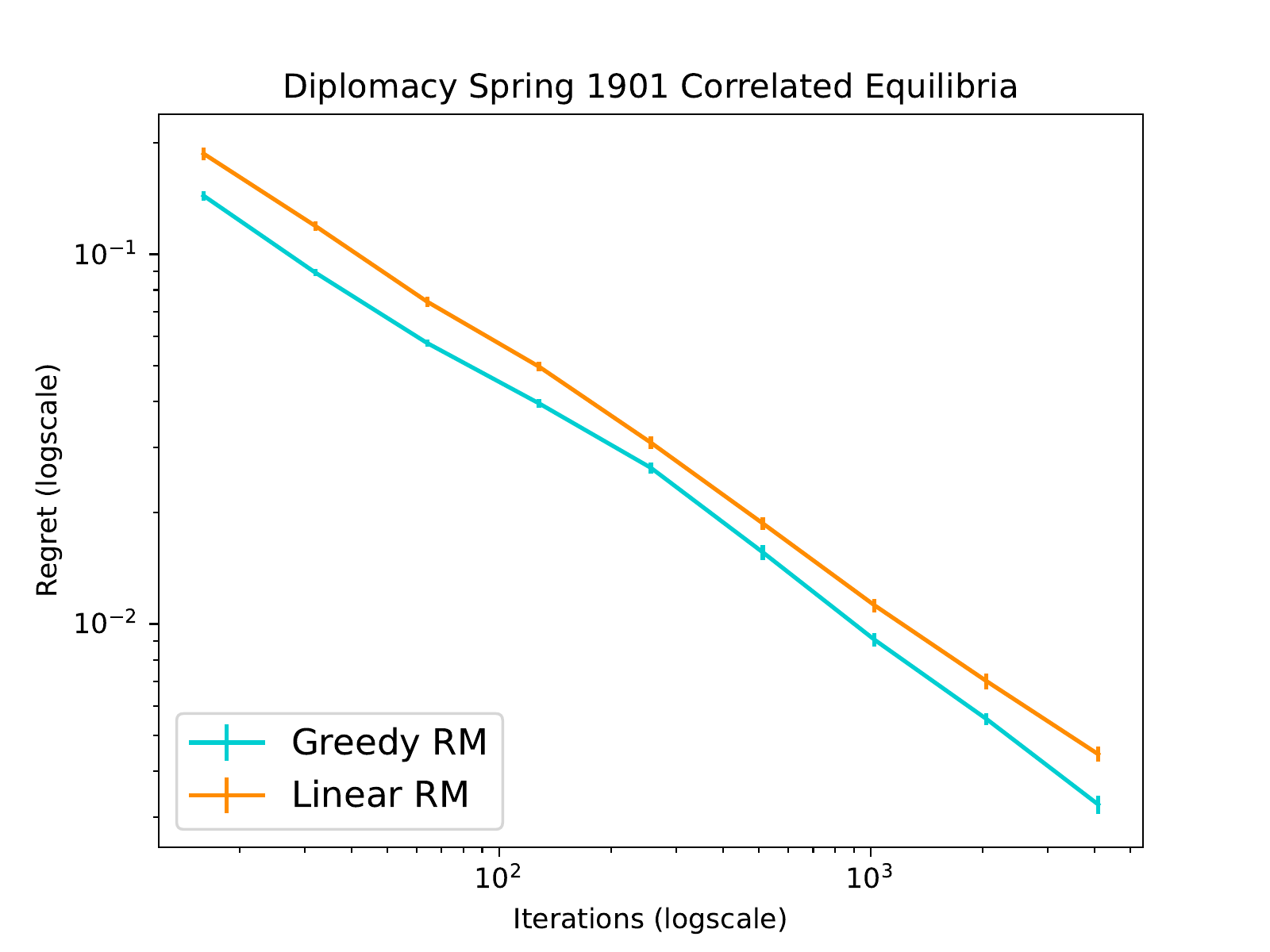}
\end{subfigure}%
\caption{We benchmark greedy weights for computing both CCE and CE on the first turn of the seven-player game of Diplomacy
which are computed by minimizing external and internal regret respectively. Greedy weights is significantly faster than in all cases considered, and when computing CCE, we find gains of several orders of magnitude. Note that both axes are logscale. Error bars denote $95\%$ confidence intervals. Additional plots for Diplomacy can be found in the Appendix.}
\label{fig:diplomacyexperiments7P}
\vspace{-10pt}
\end{figure*}

\begin{figure*}[!thb]
\centering
\begin{subfigure}[h!]{.5\textwidth}
  \centering
 \includegraphics[width=\columnwidth]{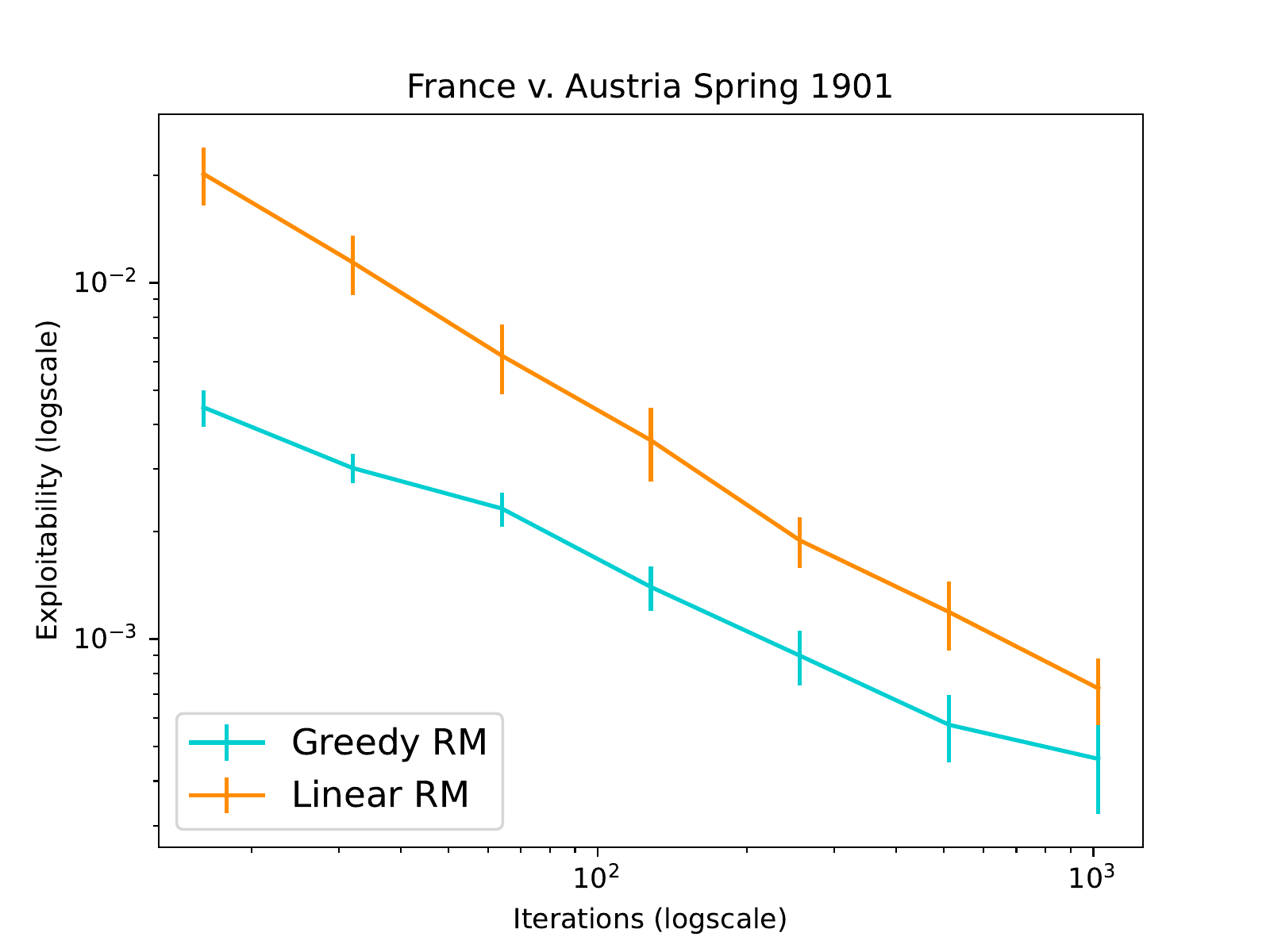}
 \caption{Spring 1901 FvA}
 \label{fig:spring_fva}
\end{subfigure}%
\begin{subfigure}[h!]{.5\textwidth}
  \centering
  \includegraphics[width=\columnwidth]{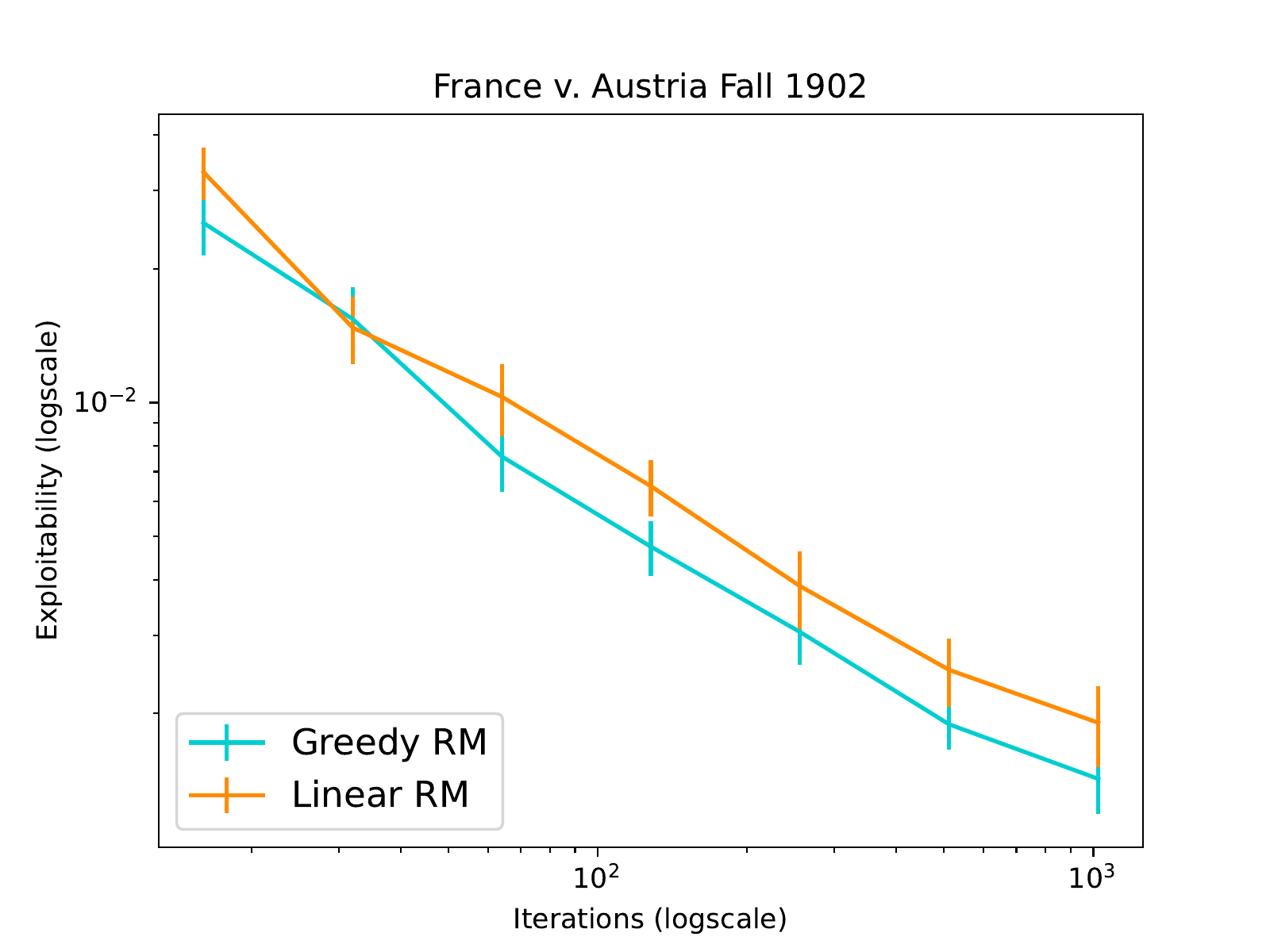}
  \caption{Fall 1902 FvA}
  \label{fig:fall_fva}
  \end{subfigure}
  \caption{We also benchmark greedy weights on a two-player zero-sum variant of diplomacy called France v. Austria. Greedy weights converges 2-3x faster than Linear RM for computing Nash equilibria in this setting. Exploitability is the $\epsilon$ of an $\epsilon$-Nash equilibrium. Note that the axes are both logscale. Error bars are at $95\%$ confidence.}
\label{fig:diplomacyexperimentsFvA}
\end{figure*}

Prior work on no-press Diplomacy (the variant where players are unable to communicate) has achieved human-level performance by using regret matching on each turn to approximate equilibrium play for just the current turn~\cite{gray2021human}. In other words, each turn is viewed as a normal-form game in which the payoffs to each player for each joint action are determined via a pre-trained value network. Since queries to this value network are relatively expensive and the number of players in Diplomacy is large, mixed-policy techniques that do not use sampling are intractable.

We run experiments on the computation of correlated equilibria and coarse correlated equilibria in Diplomacy. Additionally, we measure convergence to a Nash equilibrium in a two-player zero-sum variant of Diplomacy called France vs. Austria (FvA). In all of our Diplomacy experiments, each player chooses between the 10 actions that have highest probability in the publicly available policy network from~\citep{gray2021human}. We use the value network from~\citep{gray2021human} to determine payoffs.

Our results in Figure~\ref{fig:diplomacyexperiments7P} indicate faster convergence to both a coarse-correlated and a correlated equilibrium when compared to Linear RM. In the case of CCE, greedy weights is orders of magnitude faster. Additionally, Figure~\ref{fig:diplomacyexperimentsFvA} demonstrates that in two-player zero-sum France vs. Austria subgames, greedy weights reaches the same level of convergence to a Nash equilibrium with 2-3x fewer iterations than Linear RM when using pure policies. Experiments in Diplomacy used a single CPU core and a single NVIDIA V100 GPU. The overhead time necessary for computing an optimal weight for greedy weights was negligible relative to the cost of querying the value neural network.

\section{Conclusions and Future Work}
\label{sec:limitations}
\vspace{-0.05in}

We introduce greedy weights, a novel generalization of the regret minimization framework for learning equilibria in which each new iteration is greedily weighed to minimize the procedure's average potential function (which is a function of all player's regret). In contrast, all prior regret minimization algorithms weighed iterates according to a fixed schedule. In randomly generated normal-form games, we demonstrate that greedy weights empirically converges to correlated equilibria several orders of magnitude faster than existing methods, as well as faster convergence to Nash in two-player zero-sum games. We also find that greedy weights tends to learn equilibria with higher social welfare than vanilla regret minimization methods in general-sum games. Finally, in the large-scale AI benchmark of Diplomacy, we show speedups in convergence to all of NE, CE and CCE compared to existing methods.

Several important directions remain for future work. In this paper we focused our evaluation of greedy weights on strategic-form games. While we evaluated greedy weights on normal-form subgames of the sequential game Diplomacy,
 a natural question is whether similar results can be obtained for general sequential games that cannot be modeled as a sequence of normal-form games by extending greedy weights to counterfactual regret minimization~\cite{zinkevich2008regret}. We show some preliminary positive results in the Appendix
for the sequential games of Kuhn and Leduc poker but describe several difficult remaining challenges in scaling to large sequential games, which we believe to be worthy of future investigation.

Another interesting direction for future work is investigating alternative dynamic weighting schemes. In this paper we describe dynamically weighing iterates by greedily minimizing the potential function. This is the first regret minimization algorithm to be introduced that dynamically adjusts the weights of iterations based on information obtained at runtime. However, in theory there are numerous other ways to dynamically weigh the iterates, such as adjusting the weights of past iterates or searching ahead to future iterates. It remains to be seen whether other algorithms that dynamically weigh iterates can lead to even better performance.

Finally, greedy weights (and regret minimization algorithms in general) cannot guarantee discovery of any \emph{particular} equilibrium. While we show that greedy weights tends to find CE in general-sum game with higher average welfare
, the question of whether regret minimization procedures can be extended to find specific desired equilibria, such as Pareto optimal equilibria or the equilibrium that optimizes the sum of player utilities, remains open for resolution.



\clearpage

\bibliography{main}
\clearpage
\appendix
\onecolumn
\section{Minimizing exploitability in zero-sum games with more than two players}
\begin{figure*}[!htb]
  \centering
  \begin{subfigure}[b]{0.45\textwidth}
    \includegraphics[width=\textwidth]{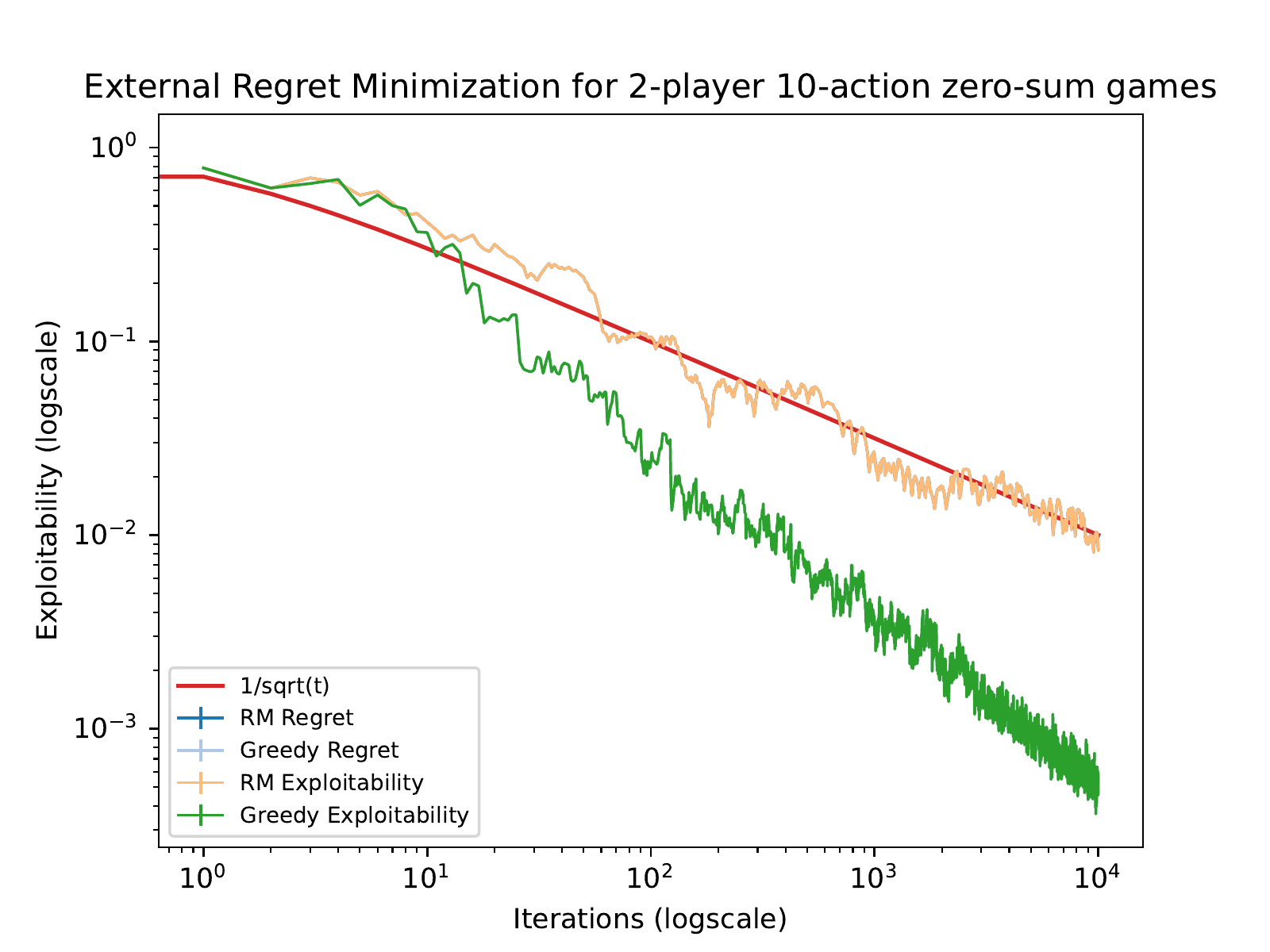}
  \end{subfigure}
  \hfill
  \begin{subfigure}[b]{0.45\textwidth}
  \captionsetup{labelformat=empty}
    \includegraphics[width=\textwidth]{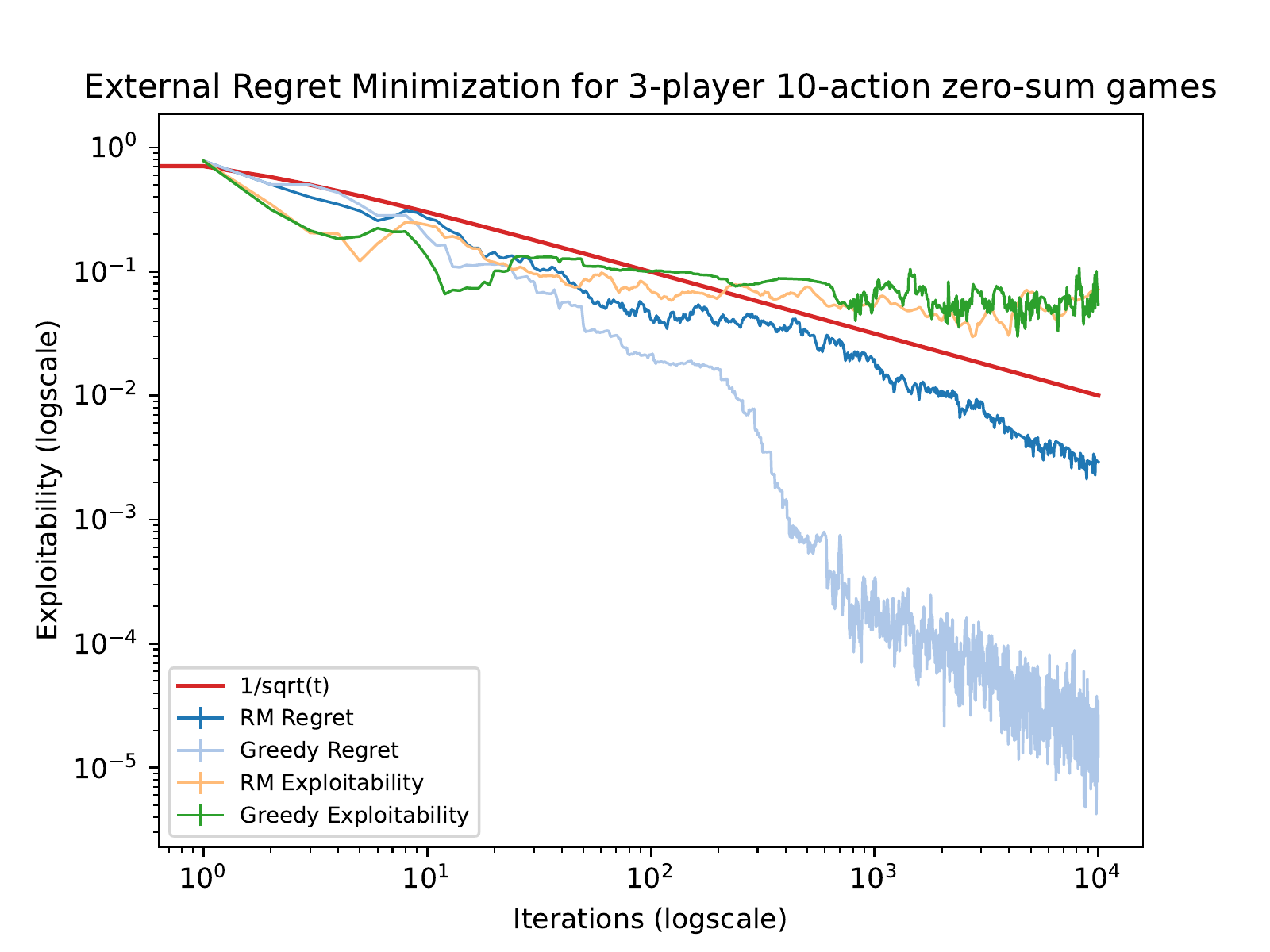}
  \end{subfigure}
  \hfill
  \begin{subfigure}[b]{0.45\textwidth}
  \captionsetup{labelformat=empty}
    \includegraphics[width=\textwidth]{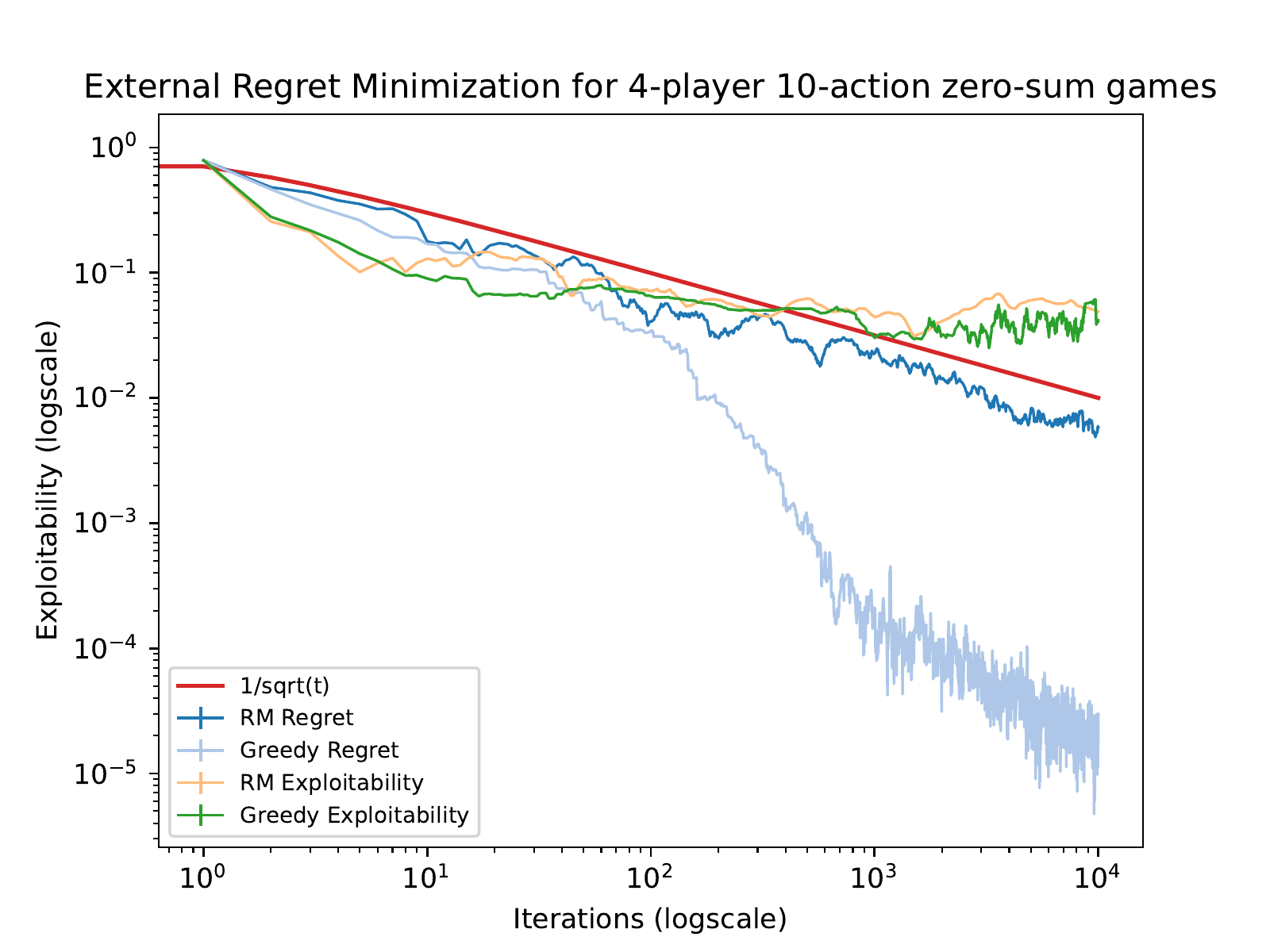}
  \end{subfigure}
  \hfill
  \begin{subfigure}[b]{0.45\textwidth}
  \captionsetup{labelformat=empty}
    \includegraphics[width=\textwidth]{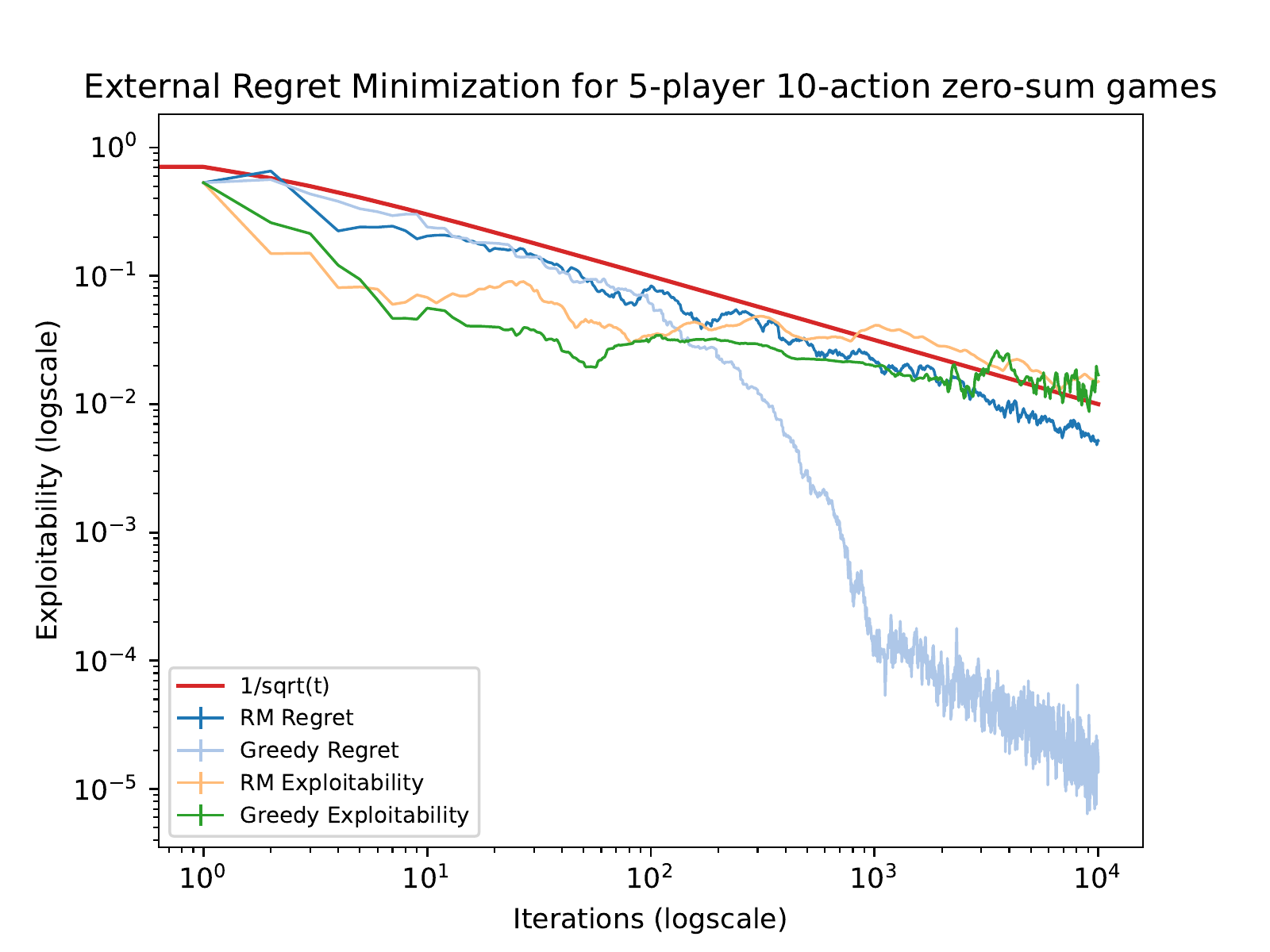}
  \end{subfigure}
  \caption{We check the relationship between (external) regret and exploitability (distance to a Nash equilibrium) in zero-sum games with between $2-4$ players. Each of these graphs depicts a single run of vanilla and greedy regret minimization in the games described. In a two-player zero-sum game (left), regret and exploitability decrease in lockstep as guaranteed by Theorem 7.2 in \citep{cesa2006prediction}. Despite the lack of theoretical guarantees for any game with three or more players, we observed that greedy weights sometimes finds an approximate Nash equilibrium with lower exploitability than vanilla regret minimization methods. Nevertheless, such gains are far smaller than the gains in accelerated regret minimization and not always guaranteed.}
  \label{fig:exploitability}
\end{figure*}

While minimizing external regret is only guaranteed to converge to the set of Nash equilibria for two-player zero-sum games, researchers have found that in practice, it often also approximates the Nash equilibria of zero-sum games with larger number of players \citep{brown2019superhuman, gray2021human}. Given that greedy weights can significantly speed up regret minimization in such games, we ask the natural question of whether it can be used to approximate the Nash equilibria of zero-sum games with 3 or more players by speeding up external regret minimization. Figure~\ref{fig:exploitability} shows our results for randomly generated games with between 2 and 5 players. In two-player zero-sum games, exploitability and regret move down in lockstep, but in games with three or more players, the two measures diverge. Nevertheless, greedy weights sometimes finds approximate Nash equilibria with lower exploitability, though the gain is not nearly as large as the gains in minimizing external regret.
\clearpage
\section{Convergence Speed As a Function of Game Size}
\begin{figure}[!h]
    \centering
    \includegraphics[width=\columnwidth]{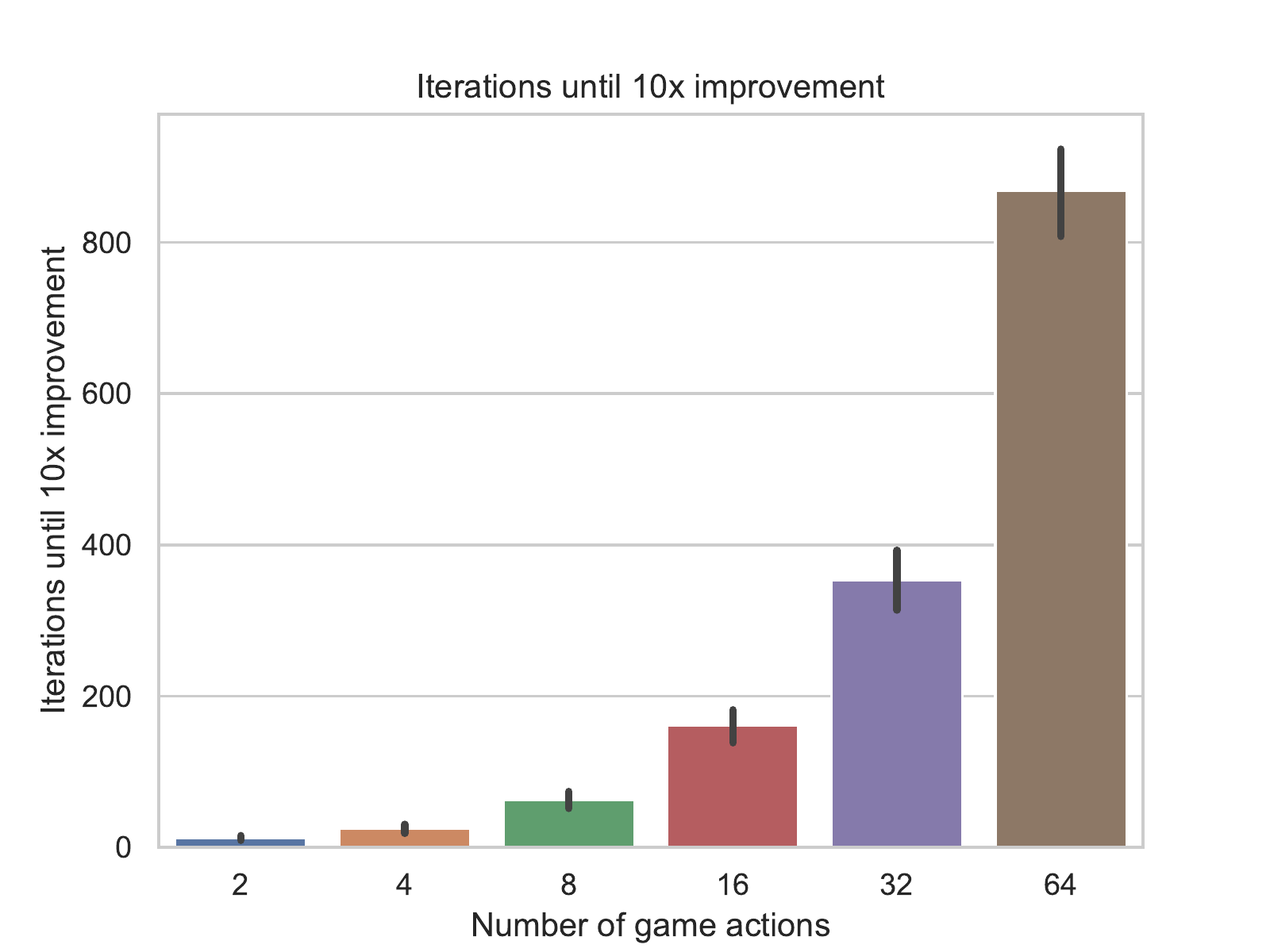}
    \vspace{-0.3in}
    \caption{Average number of iterations until 10x gain across $50$ randomly generated $3$-player games with various numbers of actions per player.}
    \vspace{-0.15in}
    \label{fig:10xgraph}
\end{figure}

In this experiment we generate general-sum games with various numbers of moves and measure the number of timesteps before greedy weights achieves regret an order of magnitude below (10x) vanilla regret minimization. Figure~\ref{fig:10xgraph} shows the average timesteps over 10 randomly generated 3-player general sum games and shows that, all else held equal, it takes greedy weights a longer time to ``pull away'' from the vanilla regret minimization as the number of moves increases.
\clearpage

\section{Choice of Objective Function}
\begin{figure}[h!]
    \centering
    \includegraphics[width=\columnwidth]{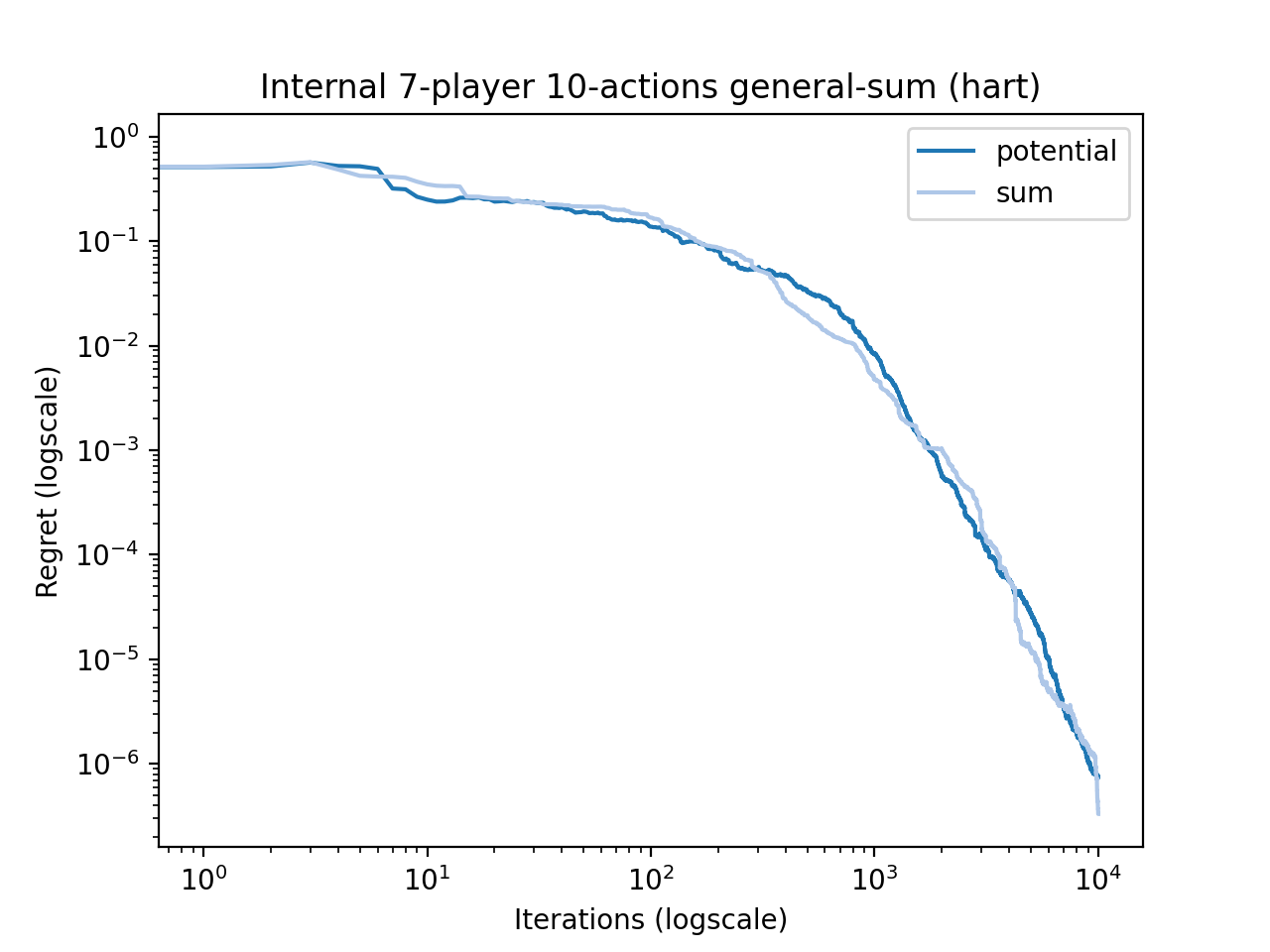}
    \caption{Minimizing the potential function performs similarly to minimizing the sum of player's regrets, suggesting that greedy weights is robust to other choices of which objective to minimize.}
    \label{fig:objective-function}
\end{figure}
For the main experiments, we minimize the sum of every player's potential function. One could also imagine that directly minimizing the sum of player's positive regrets, which directly determine the distance to an equilibrium, might be a reasonable choice. Figure~\ref{fig:objective-function} suggests that greedy weights is robust to the choice of whether to minimize the potential or the sum of regrets in randomly generated seven-player general-sum games. In our main experiments, we opt for minimizing potential because better theoretical properties which we use to prove Theorem~\ref{mainproposition}.
\clearpage
\section{Computing the Optimal Weight}
\begin{figure}[h!]
    \centering
    \includegraphics[width=\columnwidth]{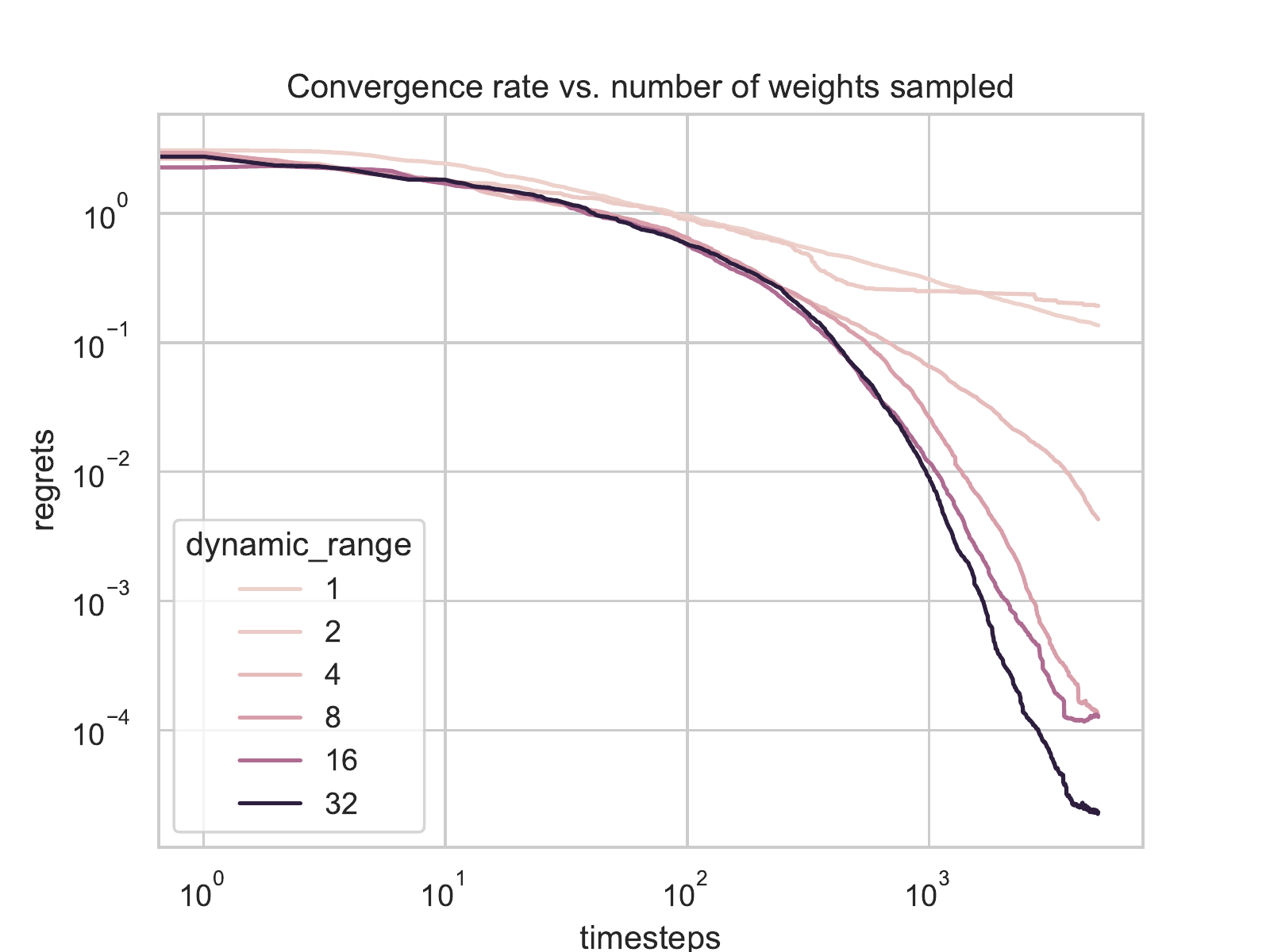}
    \caption{We measure the convergence rate as a function of the number of weights (for each iteration) considered during the greedy weights procedure by running greedy weights for $5000$ iterations across $10$ randomly generated $7$-player general-sum games. A dynamic range of $1$ is equivalent to vanilla regret minimization. This experiment suggests that adaptively selecting the weight is the primary cause of the gains observed in greedy weights rather than selecting the optimal weight.}
    \label{fig:how-many-guesses}
\end{figure}
Another natural question for greedy weights is how important choosing the optimal weight is.
Figure~\ref{fig:how-many-guesses} picks the weight for each iteration by sweeping across logarithmically spaced candidate weights between $1$ and $T^2$ instead of computing the optimal weight at each iteration. The fact that considering even a small number of guesses already substantially improves the rate of regret minimization convergence by several orders of magnitude suggests that the important aspect of our algorithm is the fact that we are greedily up or down weighting iterations based on regret, rather than finding the optimal weight.
\clearpage
\section{Greedy Weights in N-player Games}
\label{sec:greedyresultsdump}
\begin{figure}[!h]
    \centering
    \includegraphics[width=0.45\columnwidth]{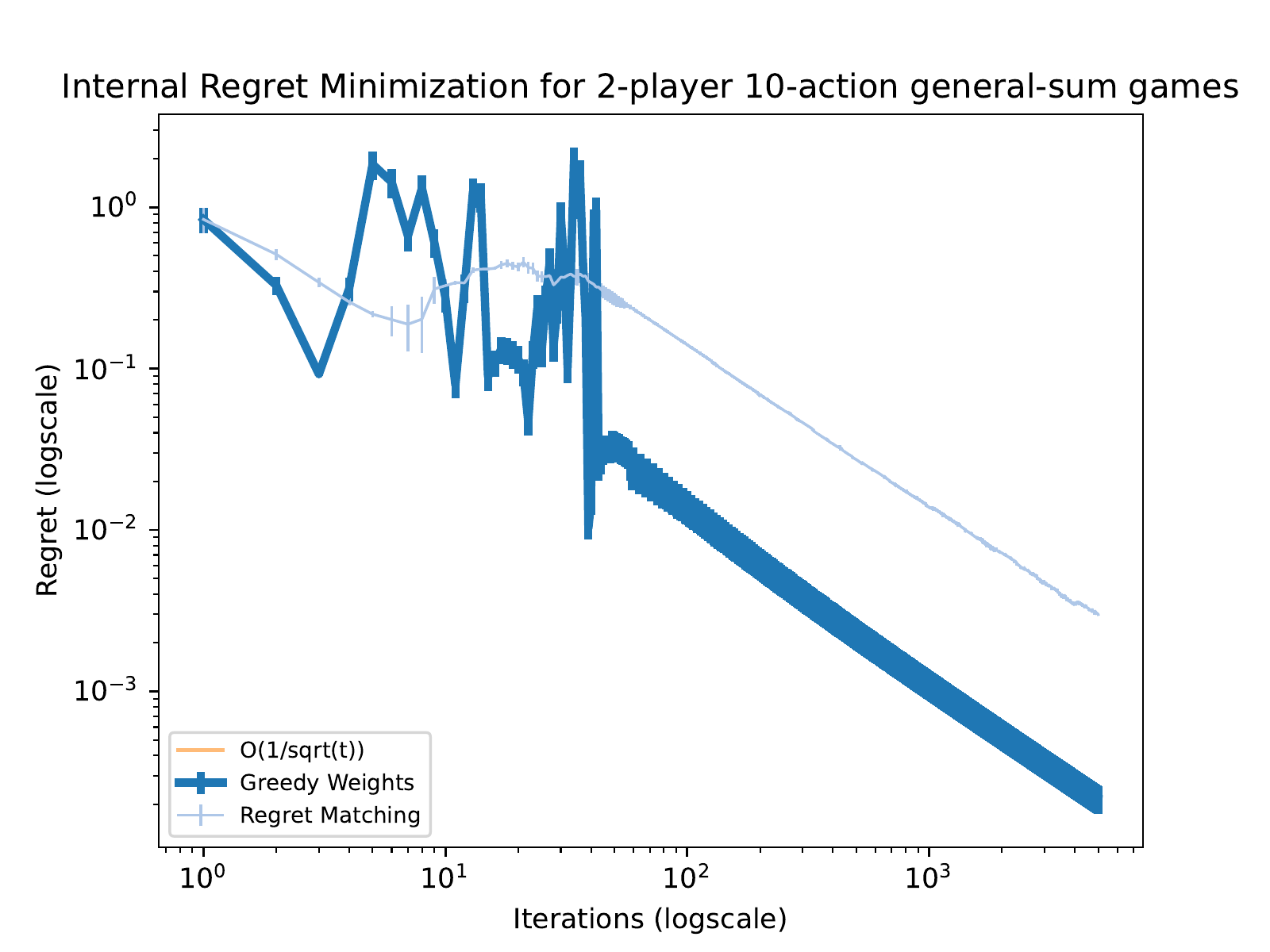}
    \includegraphics[width=0.45\columnwidth]{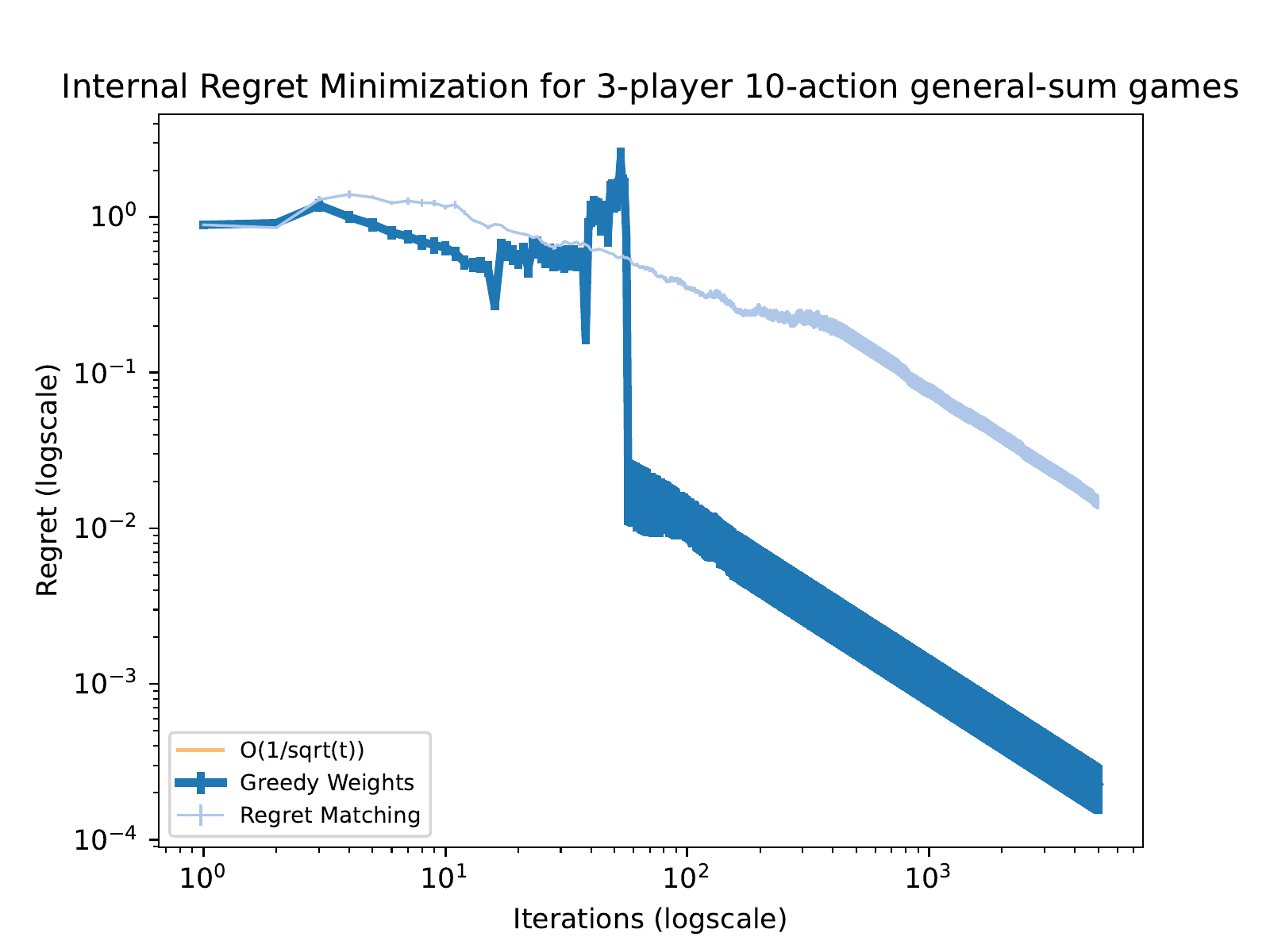}
    \includegraphics[width=0.45\columnwidth]{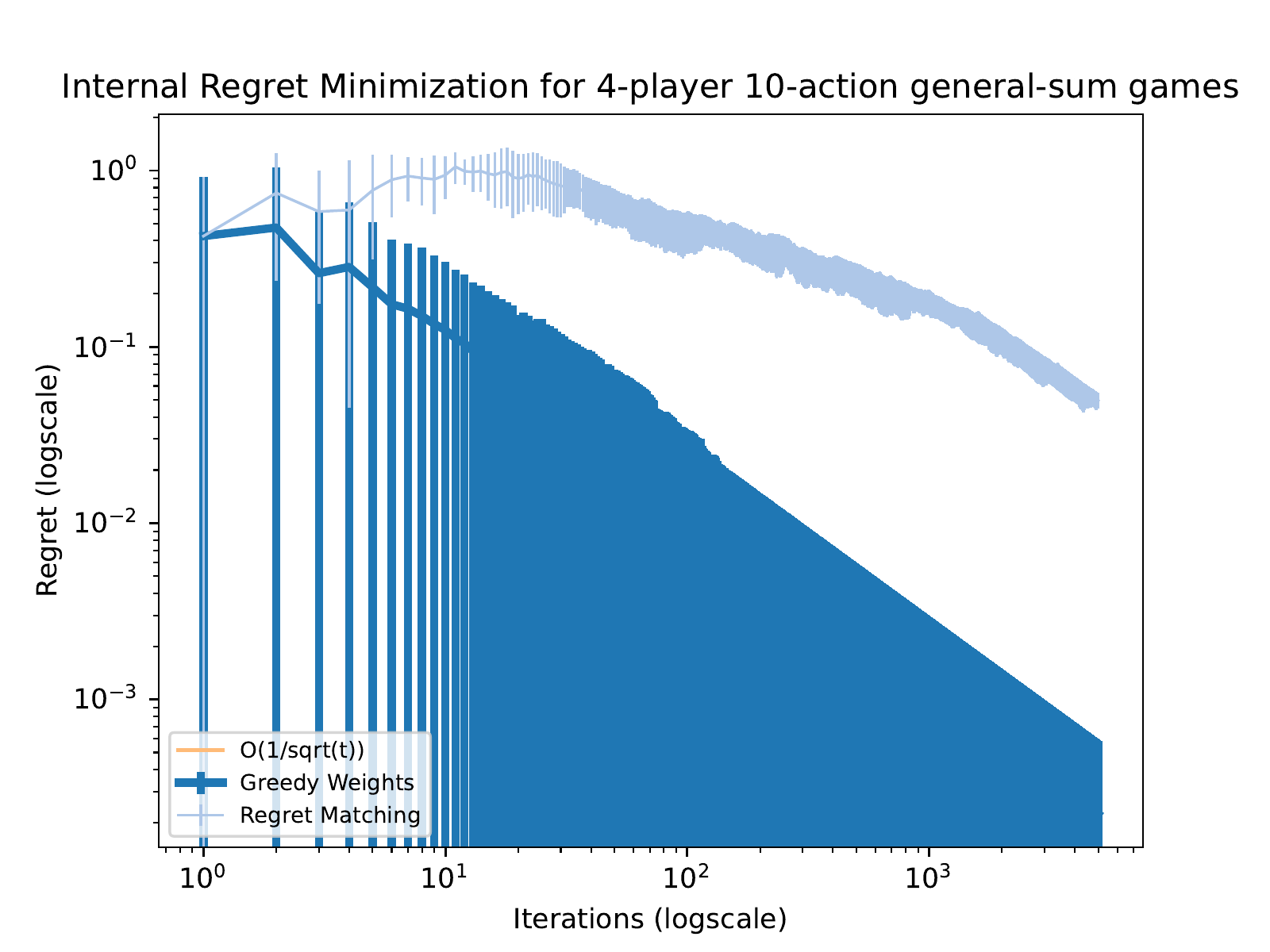}
    \includegraphics[width=0.45\columnwidth]{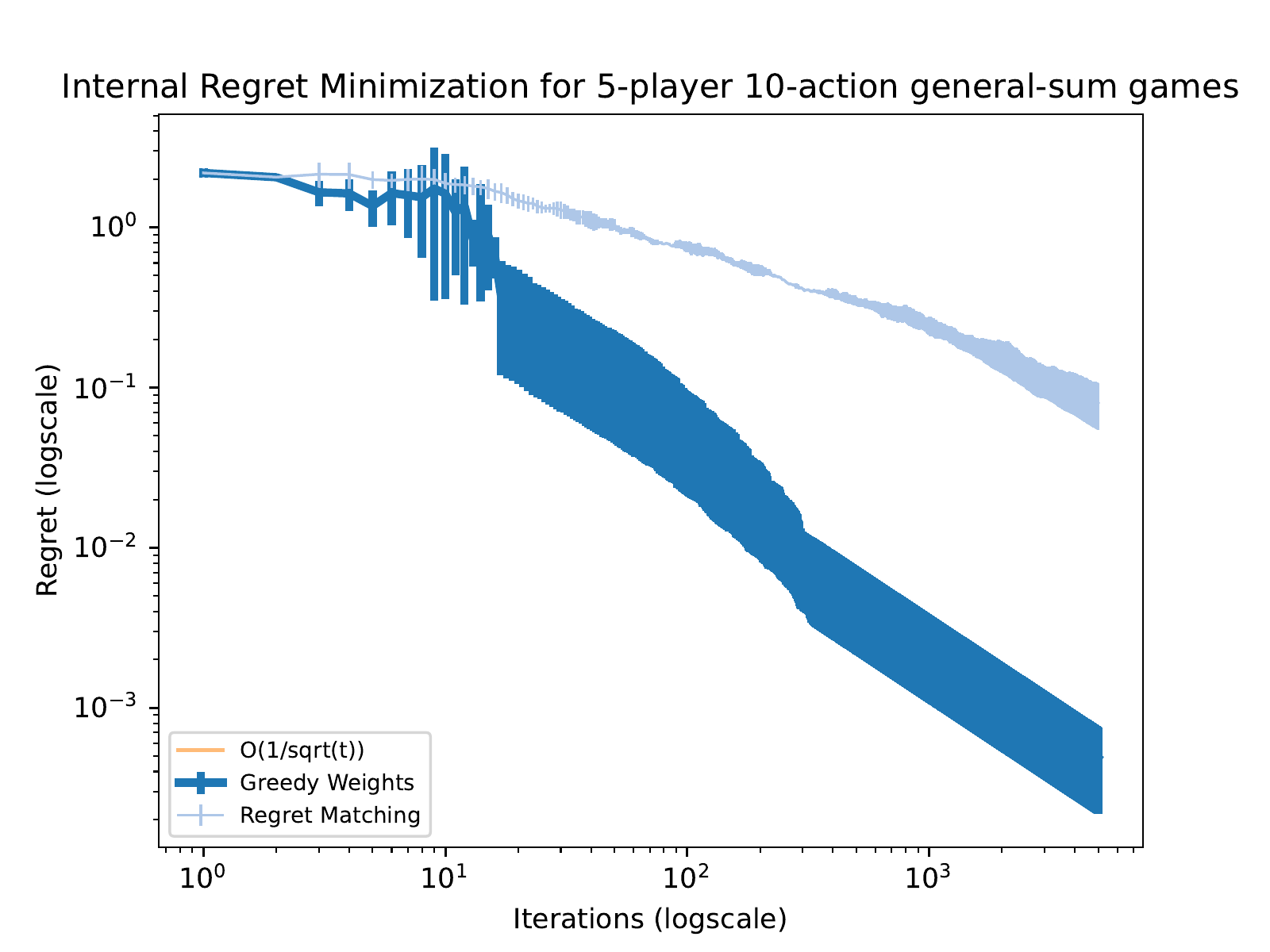}
     \includegraphics[width=0.45\columnwidth]{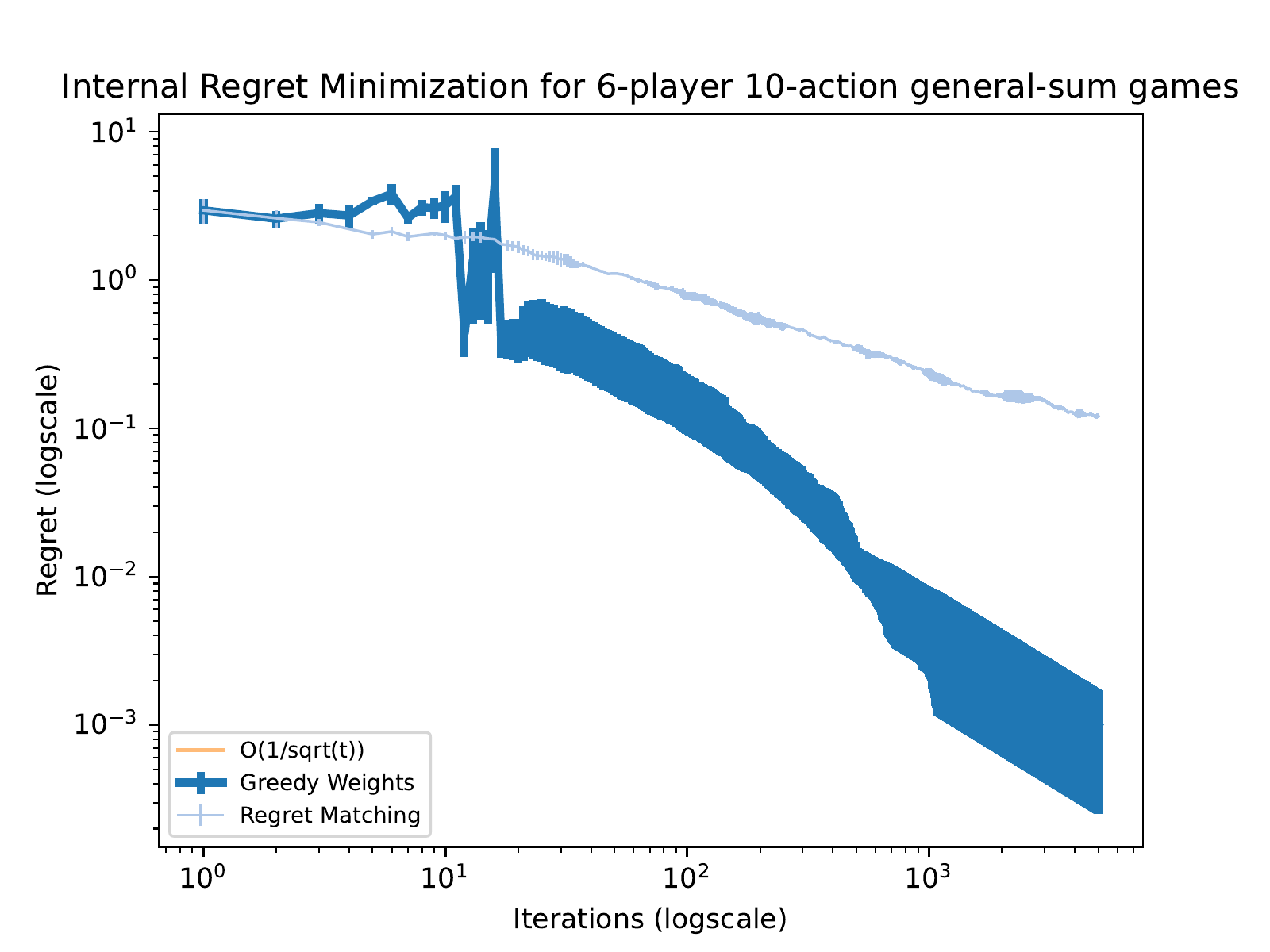}
    \caption{Greedy weights accelerates the convergence over vanilla regret matching for a variety of games. Here, we show results for randomly generated general-sum games with $2-6$ players and $10$ actions per player. Each plot is averaged over $10$ randomly generated games with different seeds and error bars at at $95\%$ confidence.}
    \label{fig:10-other-normal-form}
\end{figure}

For space reasons, we were only able to show a small portion of our full experiments with greedy weights. Figure~\ref{fig:10-other-normal-form} shows our experiments for minimizing internal regret on randomly generated general games with between $2$ and $6$ players and $10$ actions per player below. 
\clearpage
\section{Extensions to Counterfactual Regret Minimization}
\label{sec:extensive_form}
While greedy weights was designed to solve normal-form games faster, a natural question is whether it can be applied to counterfactual regret minimization techniques for solving extensive-form games without having to resort to using normal-form subgame approximations as typically done with large games like Diplomacy \cite{gray2021human}.

\begin{figure}[h!]
    \centering
    \includegraphics[width=0.49\linewidth]{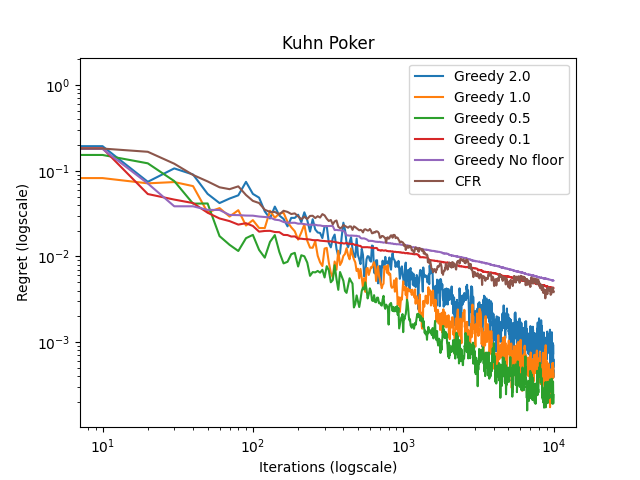}
    \includegraphics[width=0.49\linewidth]{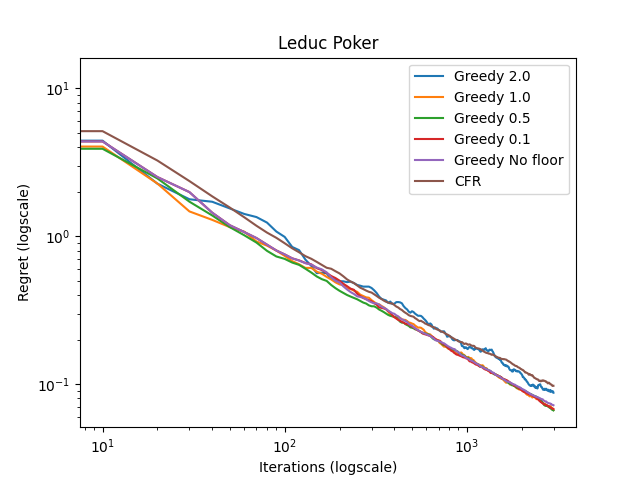}
    \caption{We test out various weight floors in both Kuhn and Leduc Poker when applying the greedy weights algorithm to counterfactual regret minimization. Consistent with the results for two-player zero-sum games in Appendix~\ref{sec:weight_floor}, we find that a relatively high floor is useful for improving convergence in such games.}
    \label{fig:efg-weightfloor}
\end{figure}

\begin{figure}[hbt!]
\centering

\includegraphics[width=0.49\linewidth]{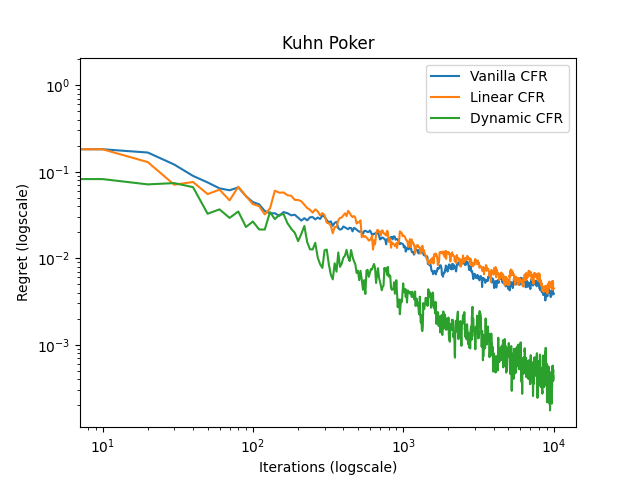}
\includegraphics[width=0.49\linewidth]{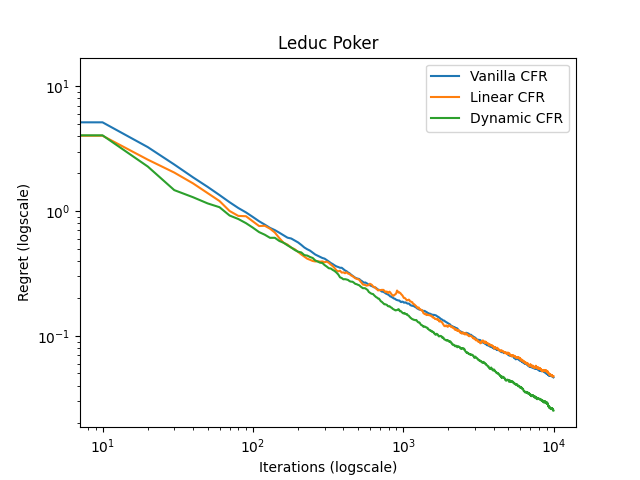}
    \caption{The left figure shows greedy weight when evaluated on Kuhn Poker. The right shows the same except for Leduc poker.}
    \label{fig:cfr}
\end{figure}

We conduct some preliminary experiments on Kuhn \cite{kuhn20169} and Leduc poker \cite{southey2012bayes} which are simplified versions of classic Texas hold'em poker via the OpenSpiel library \cite{lanctot2019openspiel}. In our experiments, the objective function being minimized is the sum of all local potential functions at all infosets. Ideally, we would like to be able to choose a different iteration weight for each infoset of the game, but retaining the theoretical guarantees of convergence rates via Theorem~\ref{mainproposition} requires that the weighting used for all infosets on a given iteration must be equal. Empirically, consistent with our normal-form experiments with two-player zero-sum games described in Appendix~\ref{sec:weight_floor}, we find that setting a relatively high weight floor works best. Experiments described in Figure~\ref{fig:cfr} were run with a minimum weight floor of $100\%$ of the average weight accrued thus far. Figure~\ref{fig:efg-weightfloor} describes the results for various floor amounts in Kuhn Poker.

While the results for Leduc and Kuhn poker depicted in Figure~\ref{fig:cfr} are promising, this extension of greedy weights to extensive-form games requires expanding out all chance nodes in stochastic games where the payoff is sampled from the distribution. Without doing so, in such a setting, greedy weights might upweight ``lucky'' iterations where the stochastic outcomes are particularly good for the players, which would lead to bias in the samples. Note that this problem is unique to dynamic weighting schemes like ours and does not apply for any fixed iterate weighting scheme like vanilla or Linear CFR where the stochastic samples eventually cancel themselves out in the limit.

State of the art solvers that scale to the full game of poker typically use MCCFR methods that often only expand out a small number of chance nodes per iteration of CFR \cite{brown2018superhuman, brown2019deep, brown2019solving}. The question of how iterates can be weighed dynamically in such a setting without incurring bias remains open, and we believe that it is a exciting direction for future research.

\clearpage
\section{Ablations on the Weight Floor in Normal-Form Games}
\label{sec:weight_floor}

\begin{figure}[h!]
\centering
\begin{minipage}{.5\textwidth}
  \centering
  \includegraphics[width=\linewidth]{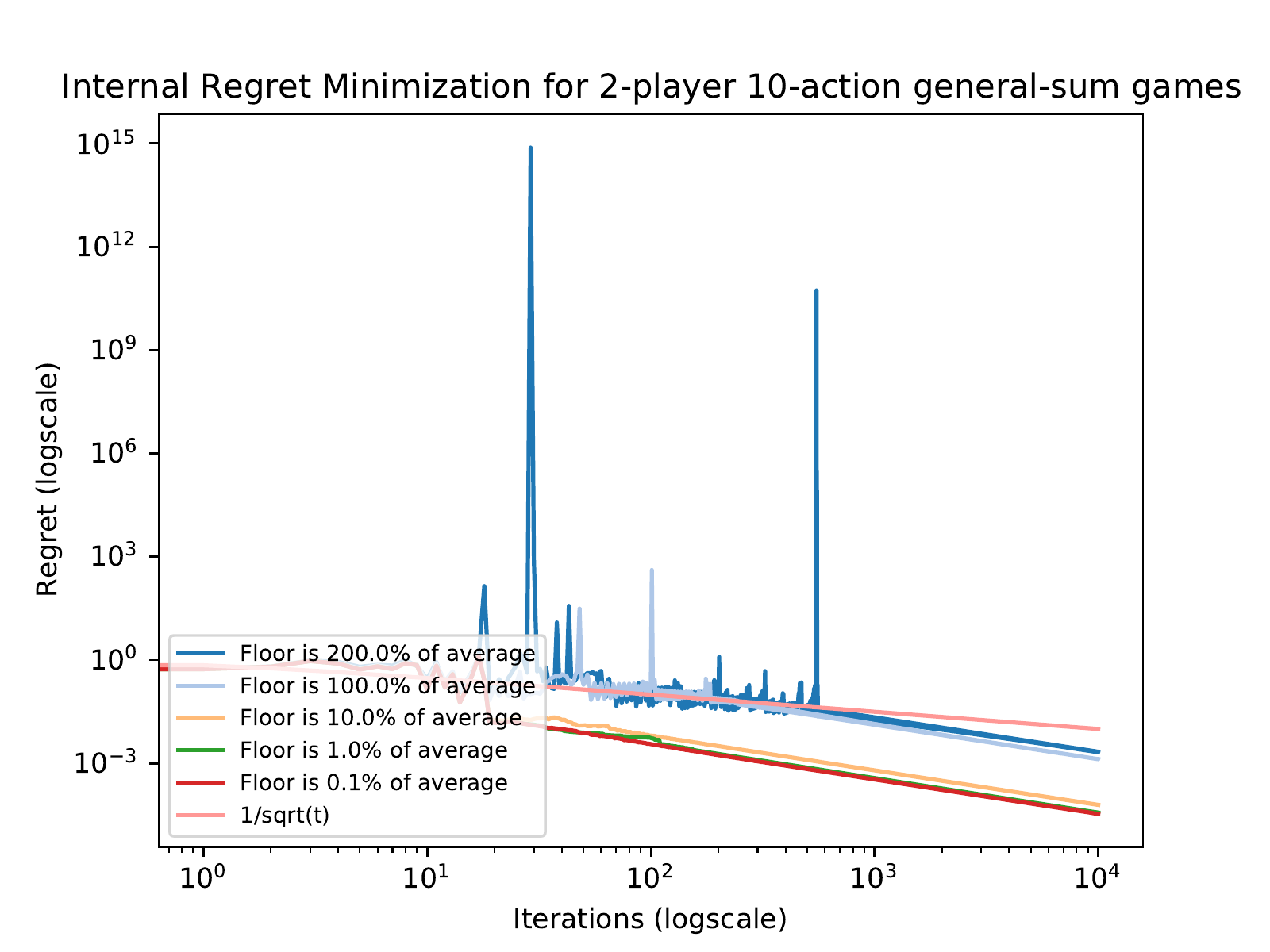}
\end{minipage}%
\begin{minipage}{.5\textwidth}
  \centering
  \includegraphics[width=\linewidth]{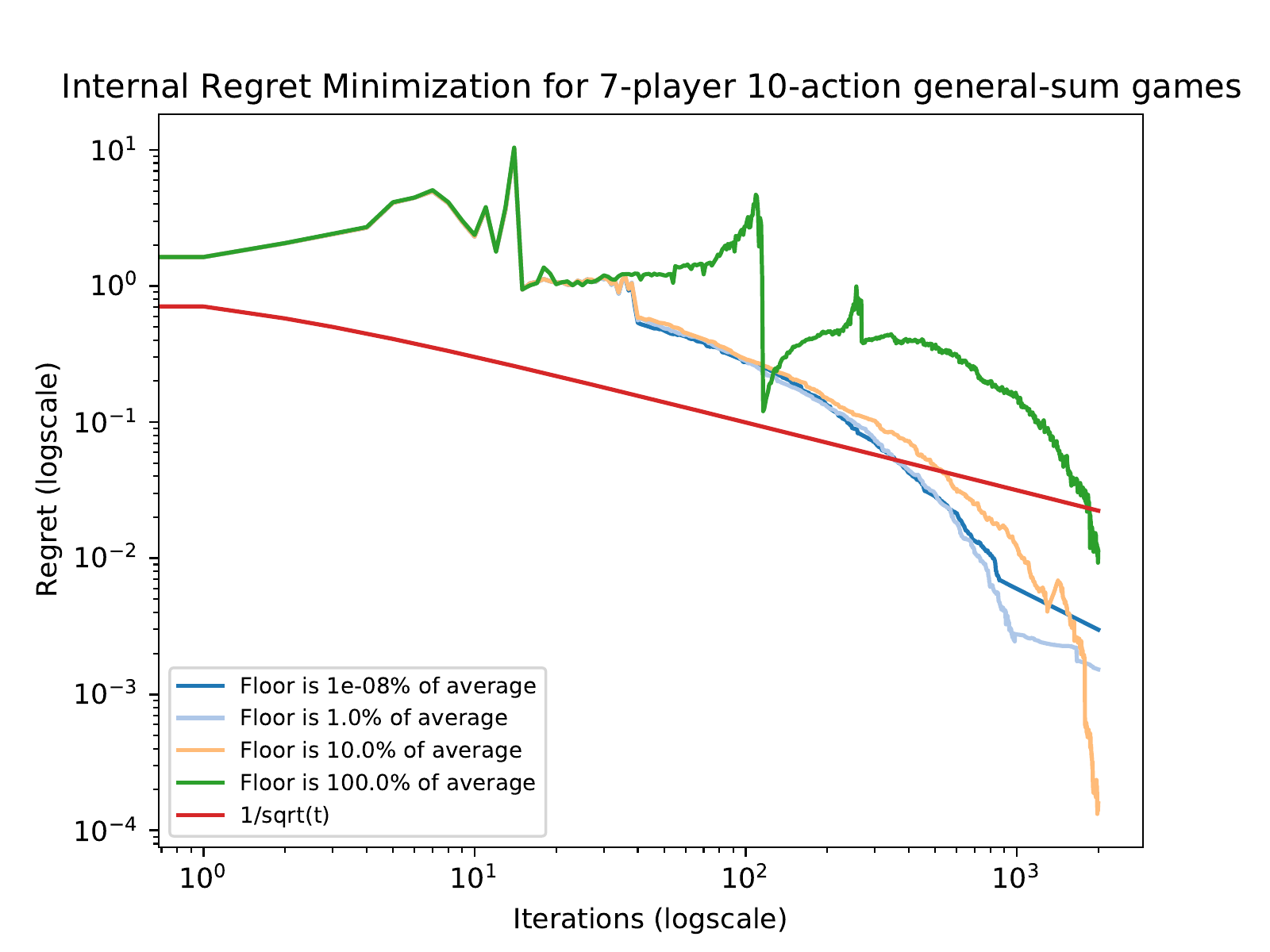}
\end{minipage}
    \caption{We examine a variety of potential weight floors for regret minimization (averaged across 10 randomly generated games) and find that large weight floors cause instability in the training procedure, while small weight floors seem to be the fastest in terms of convergence.}
    \label{fig:generalweightfloor}
\end{figure}

In Section~\ref{sec:experiments}, we suggested that in two-player zero-sum games, greedy weights benefited from a weight floor of $50\%$ of the average weight so far to improve convergence the procedure from getting stuck. Figure~\ref{fig:generalweightfloor} shows the result of several ablations on the optimal weight floor in general-sum games. In general-sum games, we find that weight floors that are too high cause unstable convergence, as greedy weights is no longer able to discard ``bad'' iterations from the regret minimization procedure.
In contrast, we find that the choice of weight floor is robust to performance differences as long as it is relatively low. As such, we choose not to use a weight floor in our main experiments with general-sum games.

In two-player zero-sum games, a weight floor appears to be beneficial when computing a minimax equilibrium (i.e., minimizing external regret). When minimizing internal regret, we find that a small weight floor is preferred, consistent with the results for general-sum games in Figure~\ref{fig:generalweightfloor}. However, when minimizing external regret, we find that a somewhat higher weight floor of 50\% to 100\% of the average iteration weight is helpful for achieving fast convergence.

\begin{figure}[h!]
    \centering
    \includegraphics[width=0.33\linewidth]{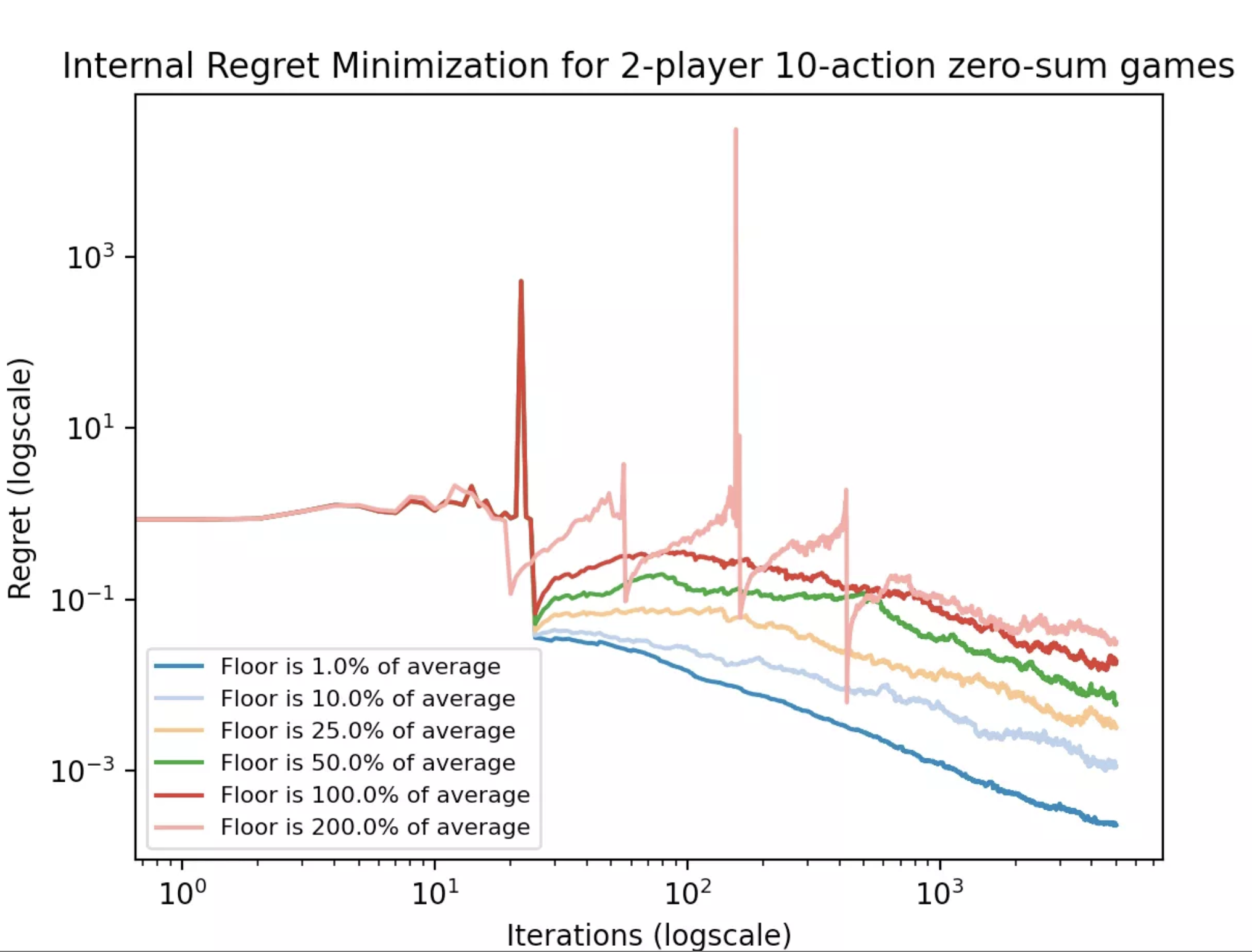}
    \includegraphics[width=0.33\linewidth]{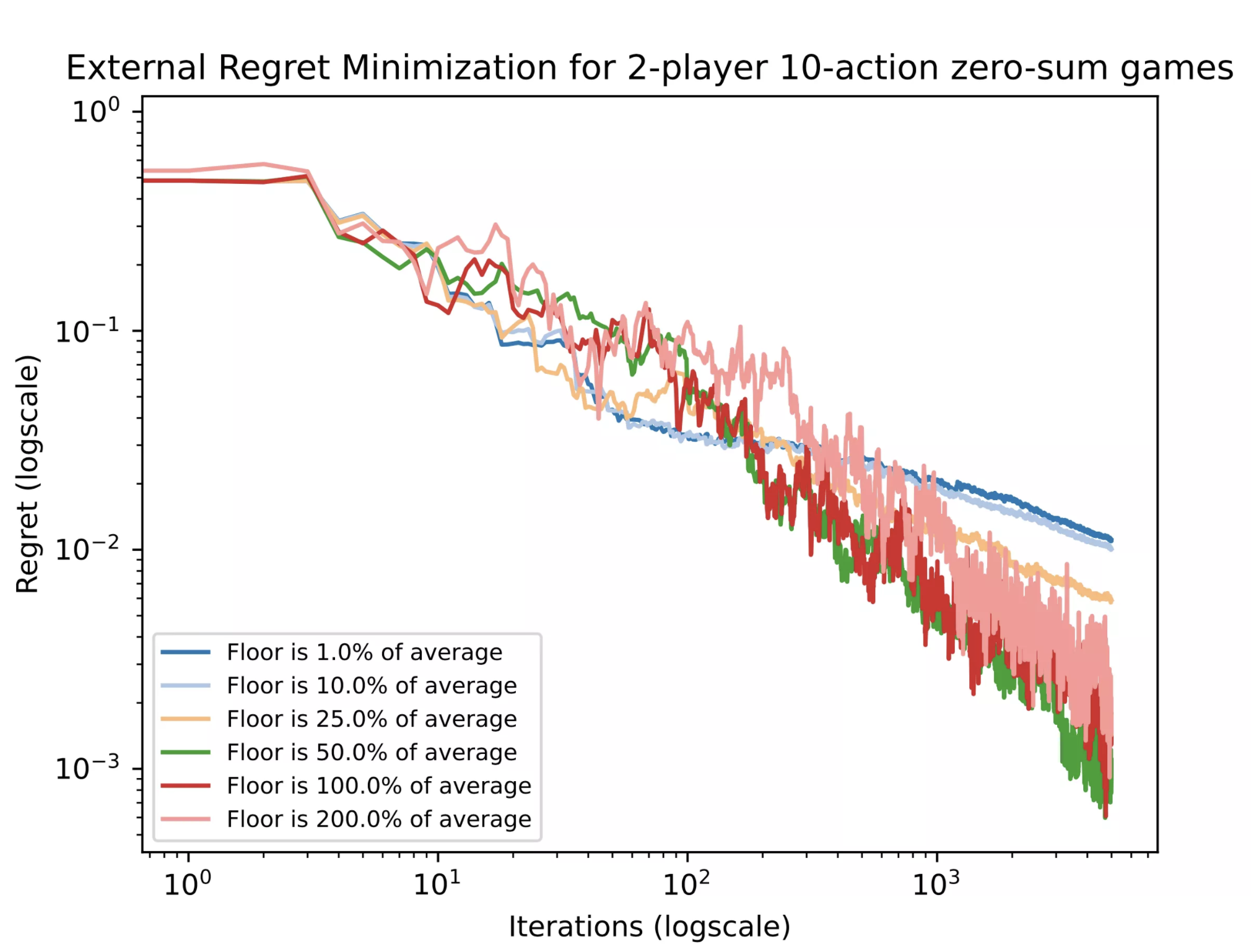}
    \includegraphics[width=0.33\linewidth]{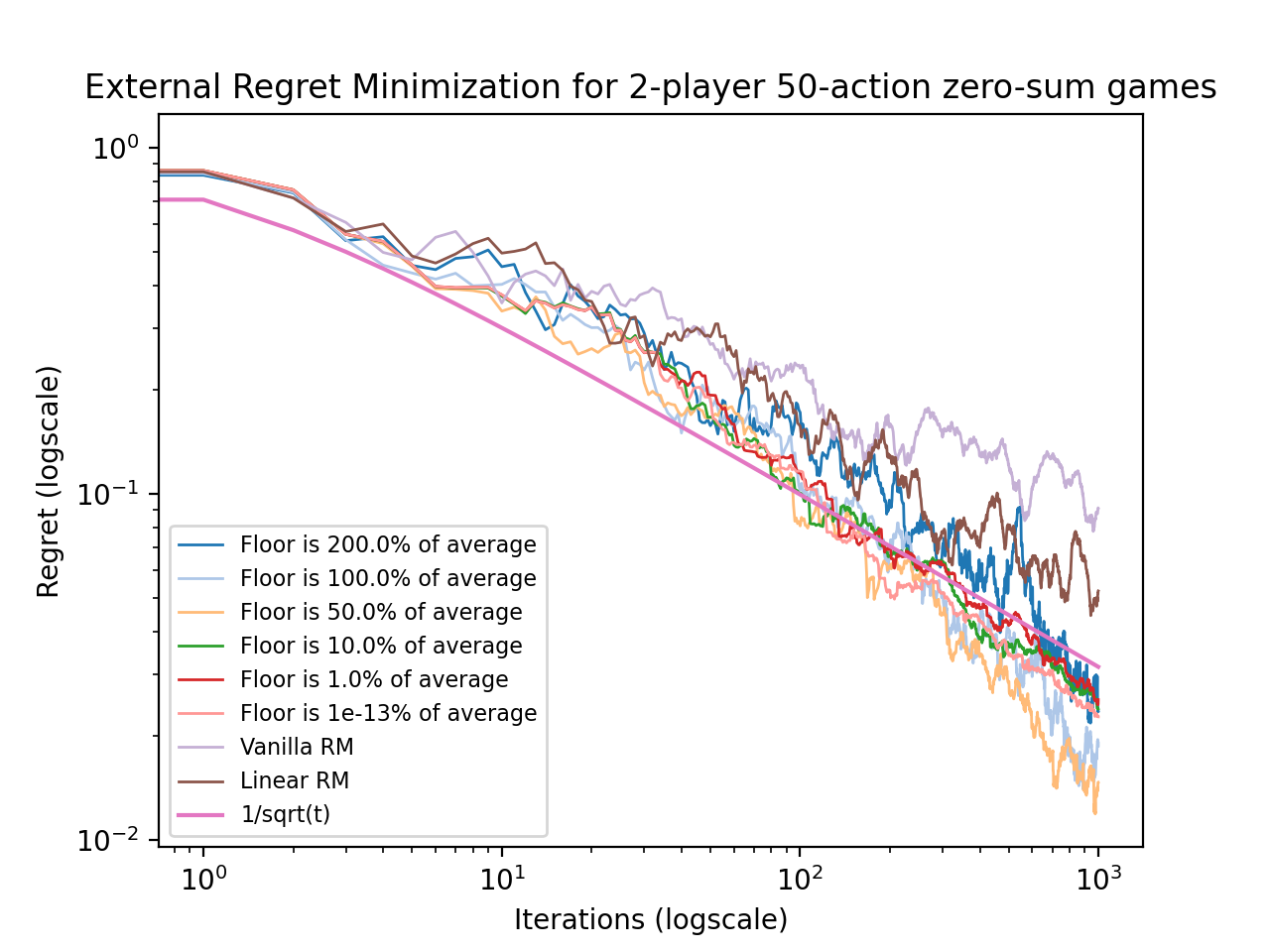}
    \caption{When minimizing internal regret (left), a low weight floor performs best, consistent with the results in Figure~\ref{fig:generalweightfloor}. However, when minimizing external regret (middle and right), we find that having a relatively high weight floor is helpful in ensuring rapid convergence, but that even in the case where no weight floor is used (right), greedy weights still performs well relative to existing methods.}
    \label{fig:zerosumweightfloor}
\end{figure}

\clearpage
\section{Ablations on Weight Floor for Diplomacy Subgames}

We also run weight floor ablations for the France v. Austria subgame and find similar results to the experiments for normal-form games described in a previous section of the Appendix. The results are depicted in Figure~\ref{fig:fvaweightfloor}.

\begin{figure}[h!]
    \centering
    \includegraphics[width=0.49\linewidth]{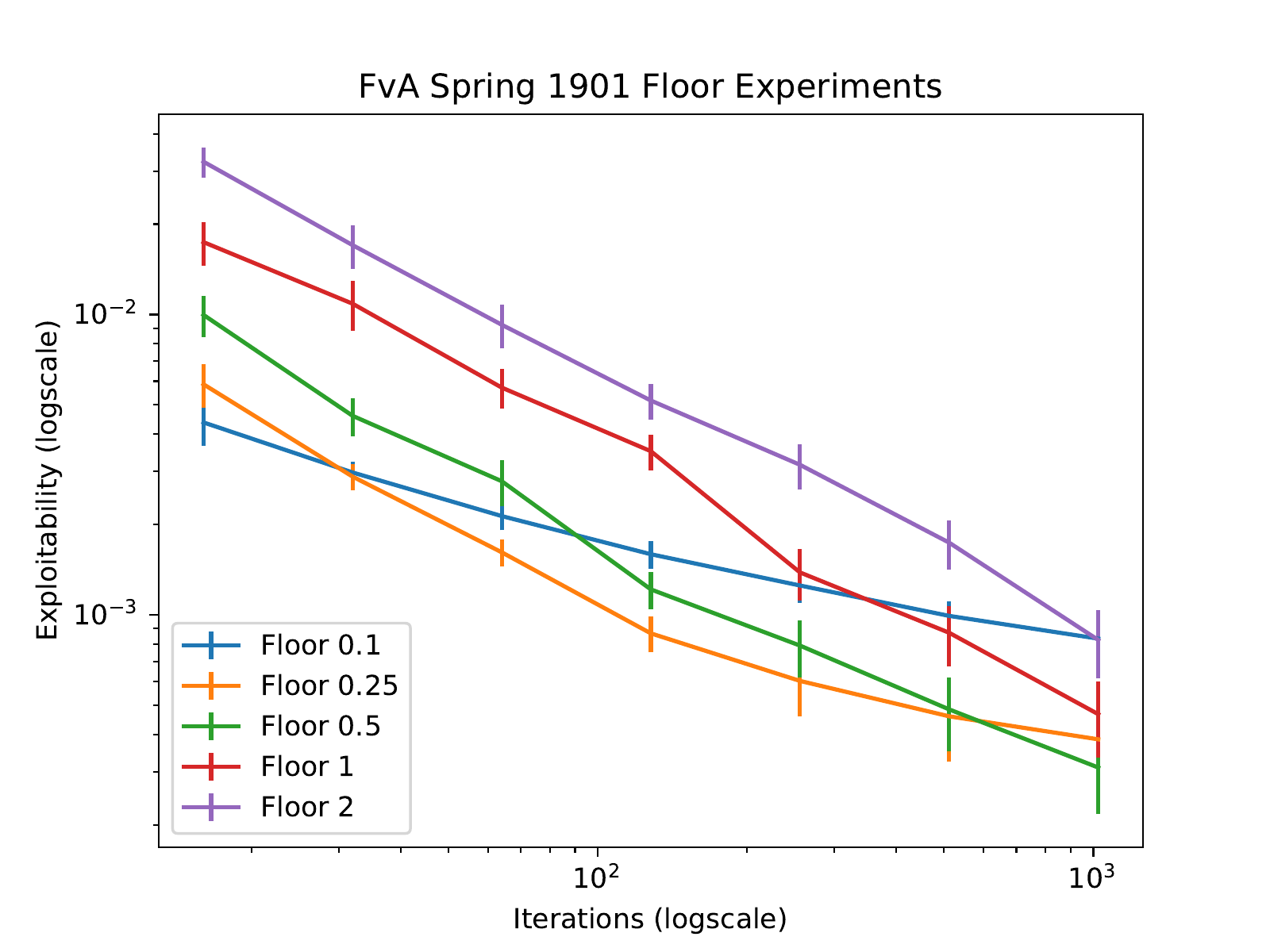}
    \includegraphics[width=0.49\linewidth]{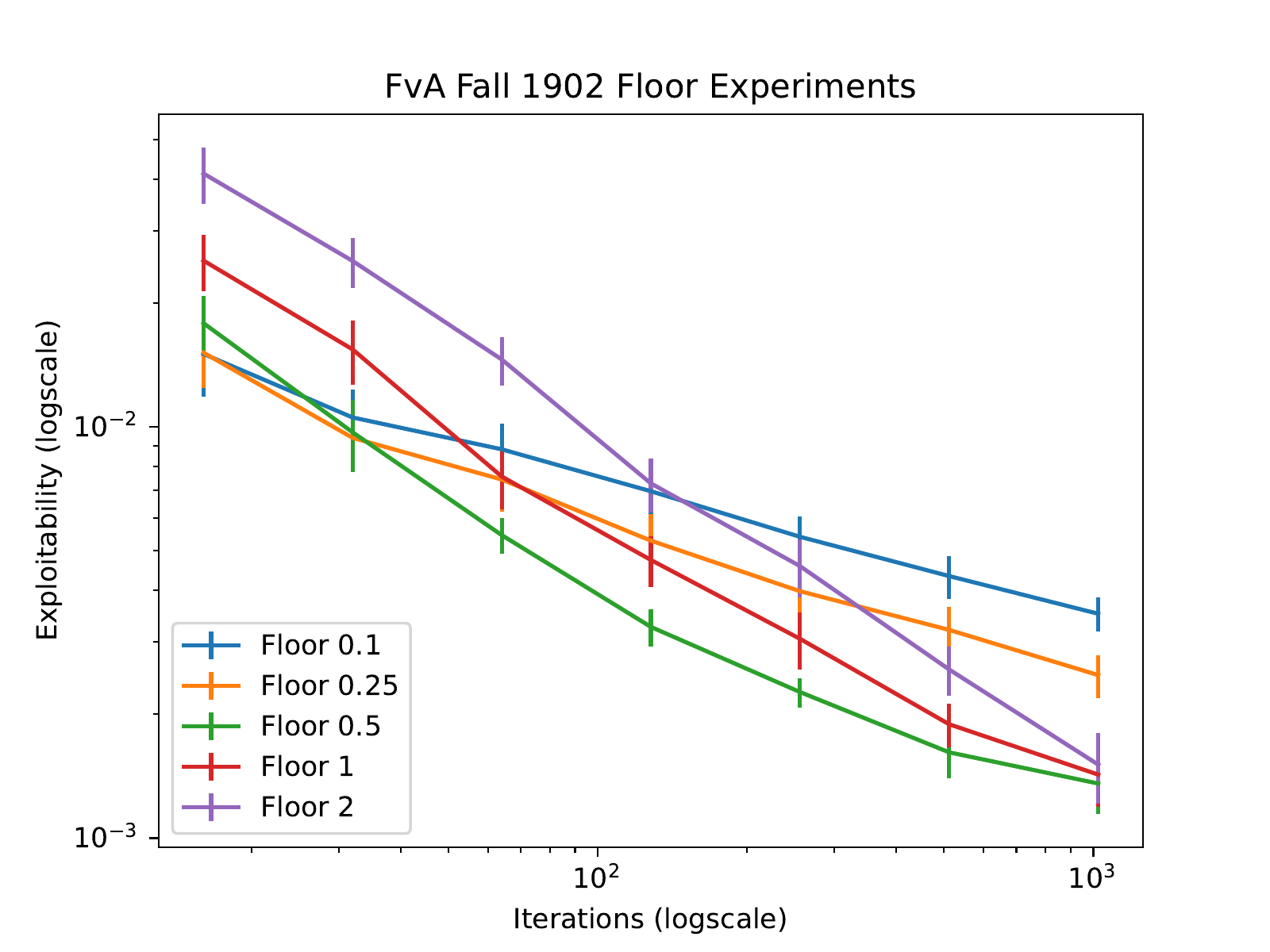}
    \caption{When minimizing external regret in France v. Austria, we find that a weight floor between $25\%$ and $100\%$ performs best. Confidence intervals at 95\%. The left figure depicts the start of the France v. Austria game, while the right figure depicts a subgame partway through the game.}
    \label{fig:fvaweightfloor}
\end{figure}

\clearpage
\section{Experiments with the Double Oracle Algorithm}

In games with an extremely large action set, one popular approach to reduce the complexity necessary to solve the game is the double oracle method \citep{mcmahan2003planning}. With the double oracle algorithm, additional actions are added one by one as needed into the regret minimization procedure instead of considering the full set of actions all at once at the start of the game.
We introduce two changes to the double oracle algorithm that we demonstrate significantly improve convergence speed. First, we use greedy weights in place of vanilla regret matching, as has typically been done throughout this paper. Secondly, we seed the first iteration of the next round of the double oracle algorithm with the approximate equilibria discovered in the latest round of regret minimization.

\begin{figure}[h!]
\centering
\begin{minipage}{.49\textwidth}
  \centering
  \includegraphics[width=\linewidth]{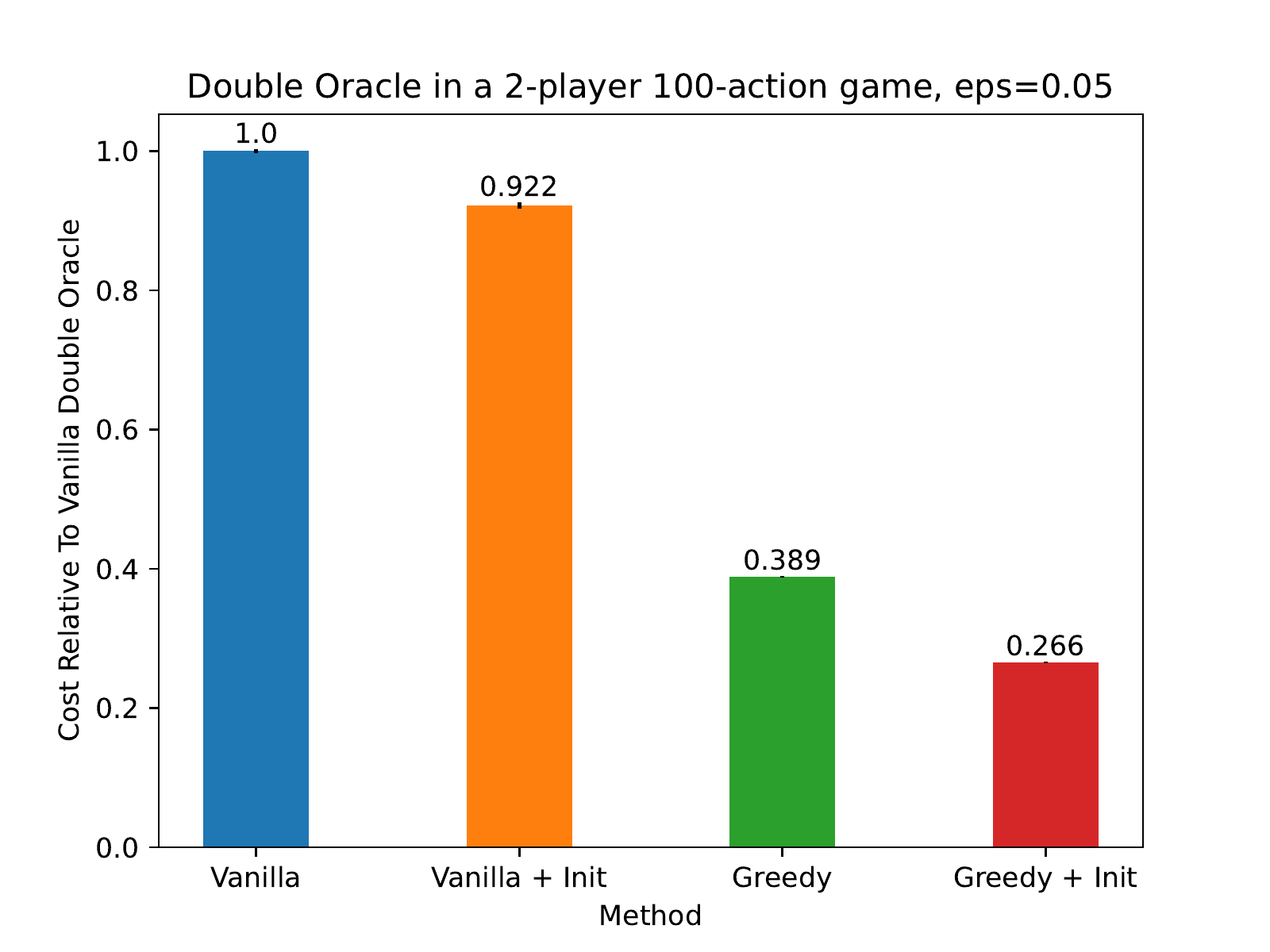}
\end{minipage}%
\begin{minipage}{.49\textwidth}
  \centering
  \includegraphics[width=\linewidth]{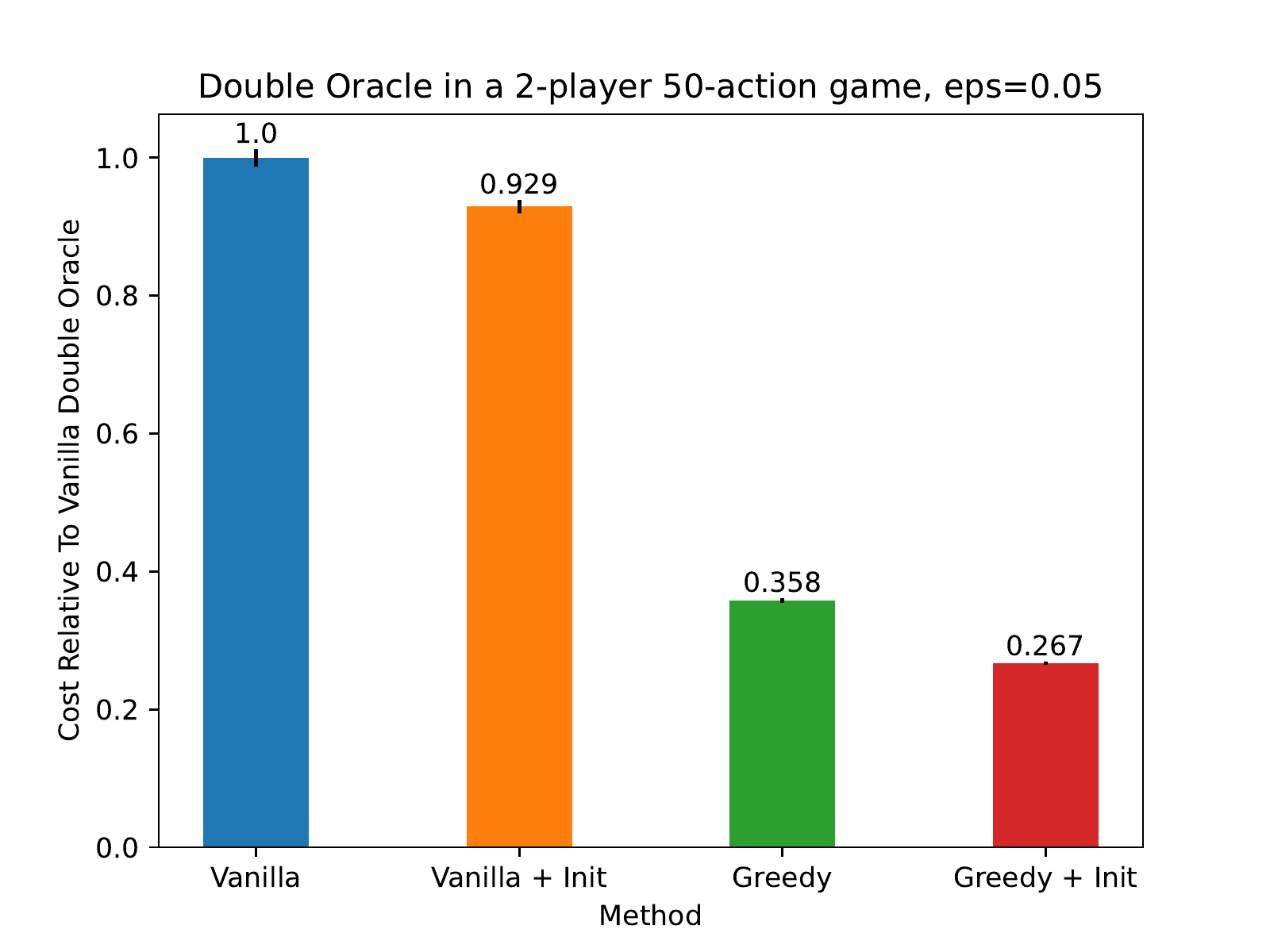}
\end{minipage}
\begin{minipage}{.49\textwidth}
  \centering
  \includegraphics[width=\linewidth]{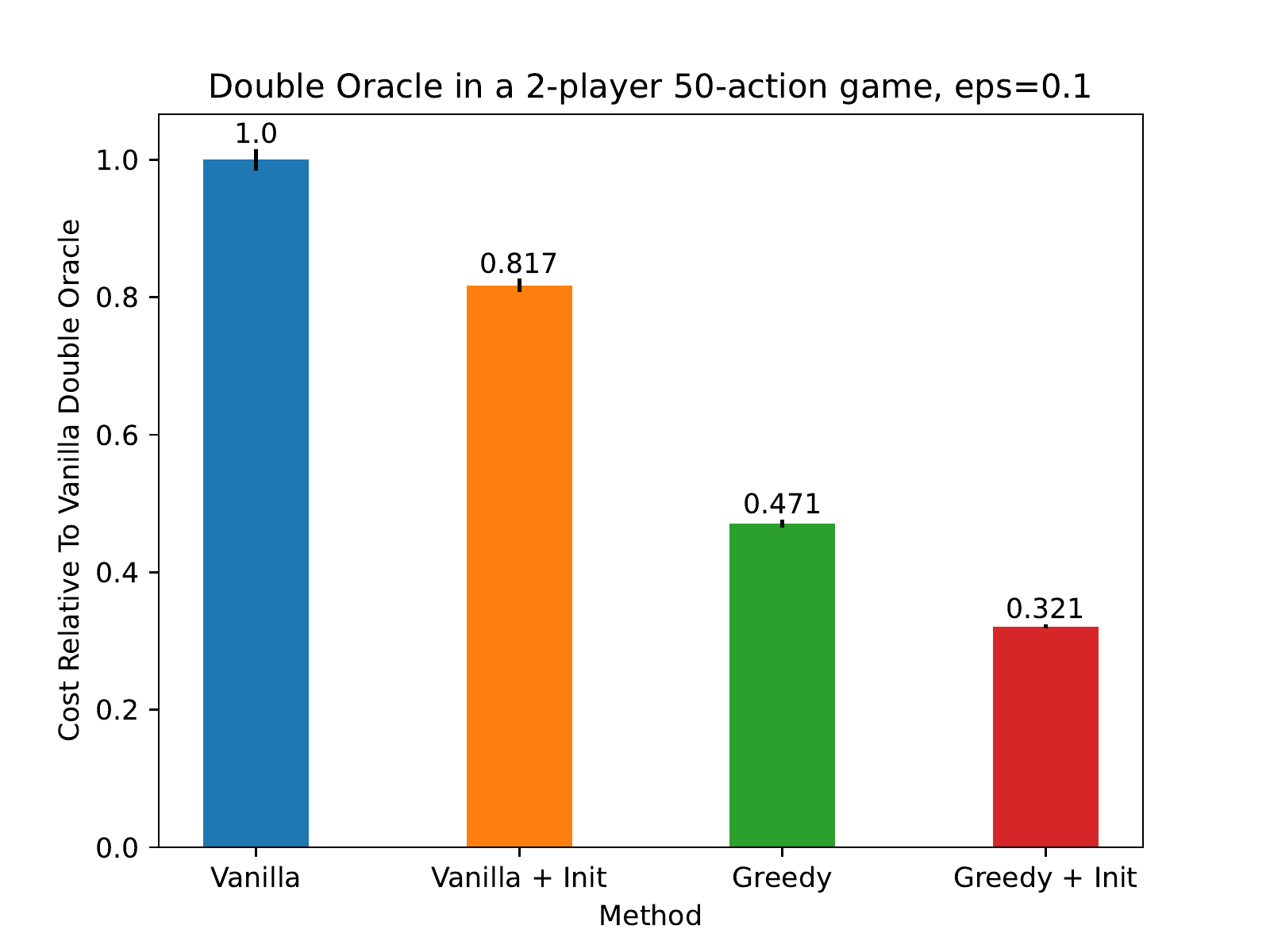}
\end{minipage}
\begin{minipage}{.49\textwidth}
  \centering
  \includegraphics[width=\linewidth]{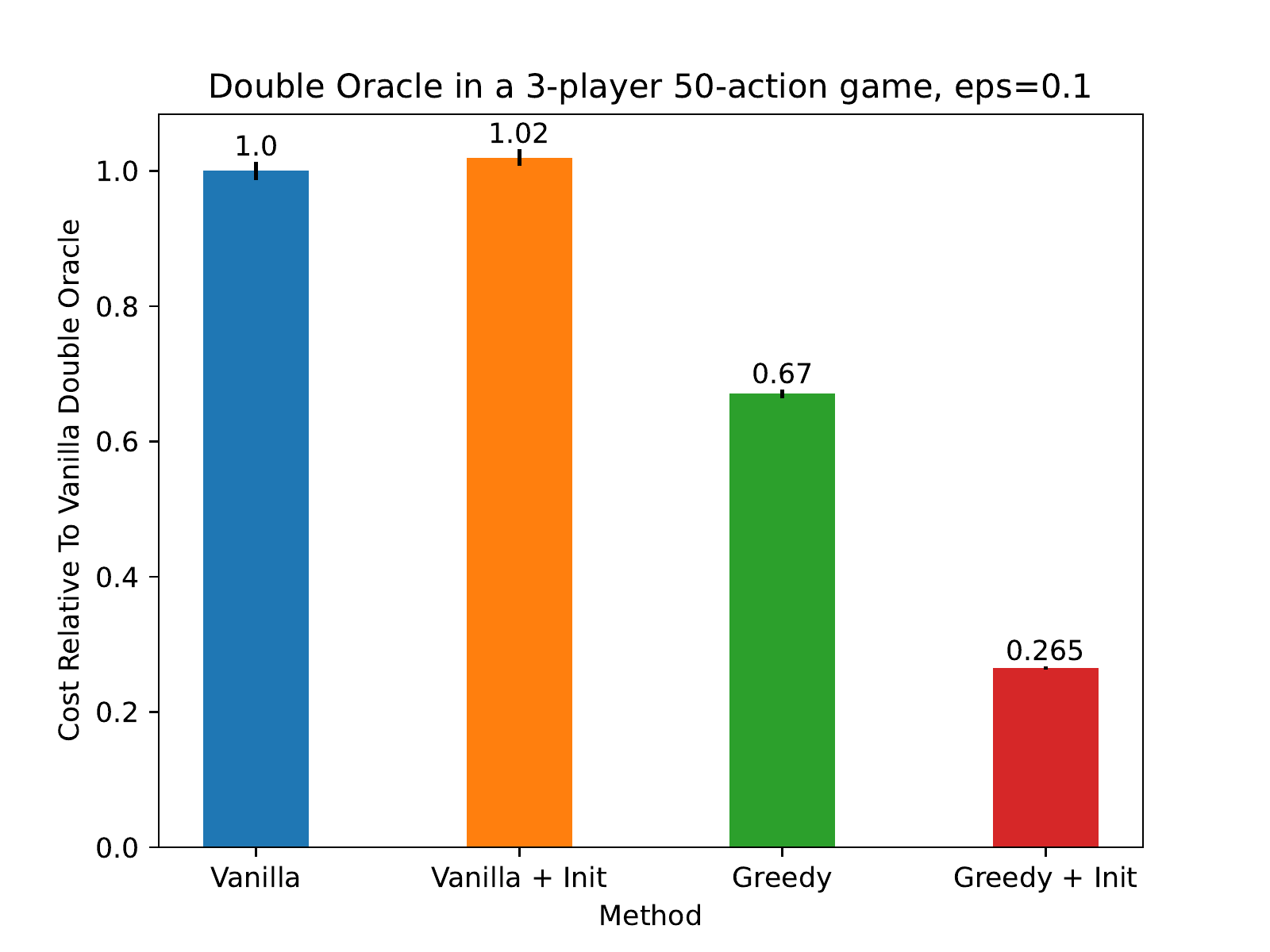}
\end{minipage}
\caption{We run double oracle and include our changes of greedy weights and seeding the previously discovered into the next iteration of double oracle. Seeding does little for vanilla regret minimization but results in substantial improvements when combined with greedy weights. $\epsilon$ measures the distance to a Nash Equilibrium. Confidence intervals are at 95\% confidence.}
\label{fig:doubleoracle}
\end{figure}

Experiments in the main paper and in Appendix~\ref{sec:greedyresultsdump} already show that greedy weights converges faster than vanilla regret minimization methods. We find that this carries over to experiments with double oracle. Furthermore, seeding the previous equilibrium discovered as the first iteration of the next regret minimization procedure provides additional gains when combined with greedy weights. This improvement is much larger than the improvement from seeding in vanilla regret matching. In vanilla regret matching, the procedure will quickly drift away from the initial moves, even if it is already very close to a Nash equilibrium, because of the fixed weighting scheme. On the other hand, greedy weights can properly weight future iterations to properly utilize seeding the regret minimization procedure closer to the equilibrium by placing lower weights on future iterations if necessary. Our empirical results are shown in Figure~\ref{fig:doubleoracle}.

\clearpage
\section{Cooperative Settings Converge Faster}

\begin{figure}[tbh!]
\centering
\begin{minipage}{.49\textwidth}
  \centering
  \includegraphics[width=\linewidth]{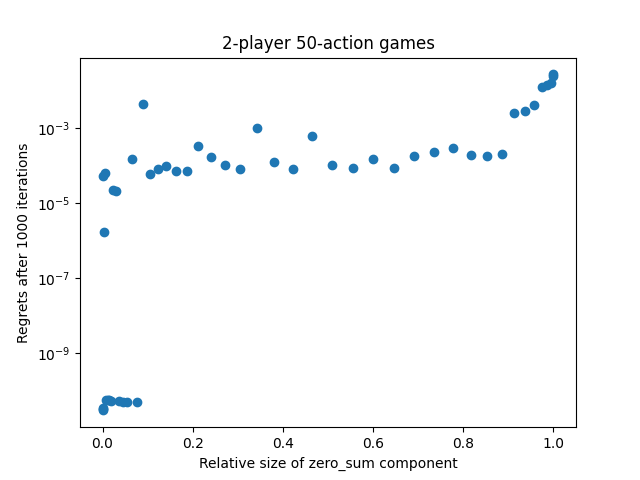}
\end{minipage}%
\begin{minipage}{.49\textwidth}
  \centering
  \includegraphics[width=\linewidth]{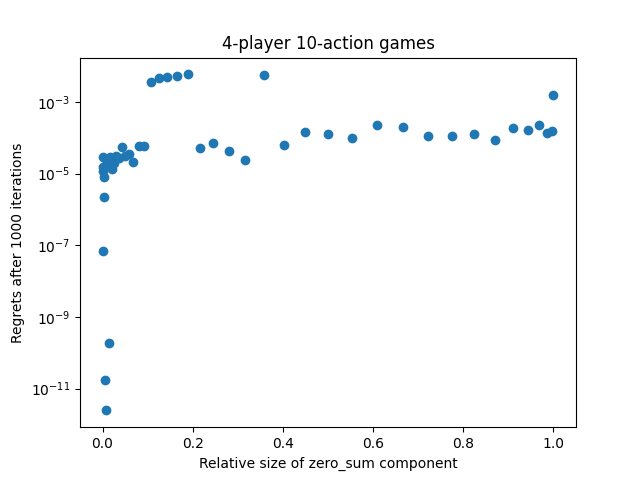}
\end{minipage}
\begin{minipage}{.49\textwidth}
  \centering
  \includegraphics[width=\linewidth]{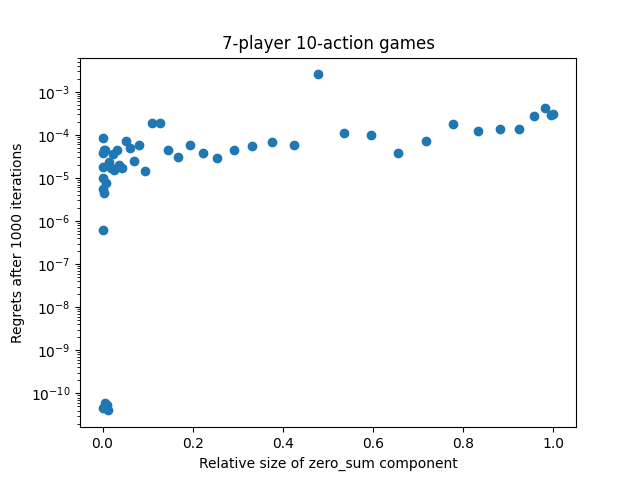}
\end{minipage}
\begin{minipage}{.49\textwidth}
  \centering
  \includegraphics[width=\linewidth]{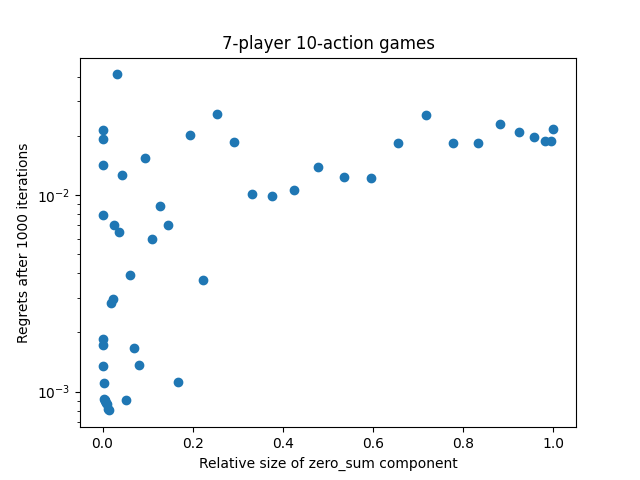}
\end{minipage}
\caption{All figures except the bottom right show regret matching with greedy weights across a variety of games. All games elicit a strong positive correlation between how adversarial the game is and how high the regret remains after $1000$ iterations of regret minimization. The bottom-right figure shows that this effect is also true, but much less strongly so, for vanilla regret matching as well.}
\label{fig:cooperative}
\end{figure}

We hypothesize a relationship between how cooperative a game is and how quickly greedy weights converges to an equilibrium. Figure~\ref{fig:cooperative} shows the result of an empirical analysis.

The game theory literature has long known \citep{kalai_engineering_2013} that the payoff matrix for every game can be uniquely decomposed into the sum of a zero-sum game and a purely cooperative game where all players share the same payoffs. Denoting $A_1 \cdots A_N$ as the payoff matrices for each of the $N$ players of the game, we can write $Z_1 \cdots Z_N$ and $C$ as the payoff matrices for the zero-sum and the cooperative game respectively.

\[
Z_n = A_n - \frac{\sum{i=1}^N A_i}{N},
C = \frac{\sum{i=1}^N A_i}{N}
\]

We measure how adversarial a game G is by performing the above decomposition and computing $\frac{\norm{Z}}{\norm{G}}$ where the norm is the Frobenius Norm of that game’s payoff matrix. A purely zero-sum game will be decomposed such that C is a zero-matrix, while a purely cooperative game will have Z be a zero-matrix, so all games will be scored between 0 and 1, where 1 indicates that the game is purely zero-sum and 0 indicates that the game is purely cooperative. We randomly generate games along the entire spectrum of this measure by generating random zero-sum and cooperative games at random via the procedure described in Section~\ref{sec:experiments} and interpolating between the two. We plot the regret after running $1000$ iterations of regret matching with greedy weights as a function of the zero-sum score of the game and find a strong positive correlation. This also applies to vanilla regret matching, though the effect is much smaller in magnitude.

Theoretical analysis of regret matching that improves the \rt bound has proved difficult, despite a wealth of empirical evidence suggesting that the bound is loose in practice. Nevertheless, there is reason to believe that this correlation is no accident. Fully cooperative games are known to always admit a pure Nash equilibrium \citep{monderer1996potential}, and thus, regret matching with greedy weights will converge asymptotically after only a constant number of iterations, since greedy weights will give infinite weight to the pure equilibrium the moment it is discovered and the size of the game is constant with respect to the number of iterations run. On the other hand, zero-sum games are more similar to the adversarial case where Theorem~\ref{mainproposition} proves that convergence can occur no faster than \rt \footnote{There are some subtle differences in that the adversary in Theorem~\ref{mainproposition} is more powerful than an opponent in a two-player zero-sum game in that she is able to pick \emph{any} regret vector for each timestep without the constraints of picking from the game matrix.}.

As a partial toy example for why vanilla regret matching is unable to take as much advantage from cooperative game settings as greedy weights, imagine a two-player game with $N$ moves where player one is a dummy and receives equal utility from all possible game states and each entry of player two’s reward matrix is drawn uniformly at random from the interval $[0, 1]$. If they both play according to vanilla regret minimization, our dummy player will choose moves randomly so the central limit theorem guarantees that the regret can proceed no faster than \rt. 

However, if greedy weights is used, the procedure will converge to an equilibrium at a worst case rate of $\frac{1}{T}$, simply because the expected number of samples before stumbling upon an entry in the interval $[1 - \epsilon, 1]$ is $O(\frac{1}{\epsilon})$. In practice, greedy weights converges even faster than this toy example because regret matching because general-sum games tend to have better than random samples for the next possible move as players learn to cooperate.

While further work must be done to explain the convergence of games that are neither purely cooperative nor purely adversarial, we suspect that further study in this direction may lead to a long sought understanding of why regret matching performs so well in practice, even though its asymptotic bound is substantially worse than other methods.




\clearpage
\section{Additional Diplomacy Plots}
We include several plots for minimizing external and internal regret for a subgame in the middle of the Diplomacy game.
\begin{figure*}[ht!]
\centering
\begin{subfigure}[b]{.45\textwidth}
  \centering
  \includegraphics[width=\columnwidth]{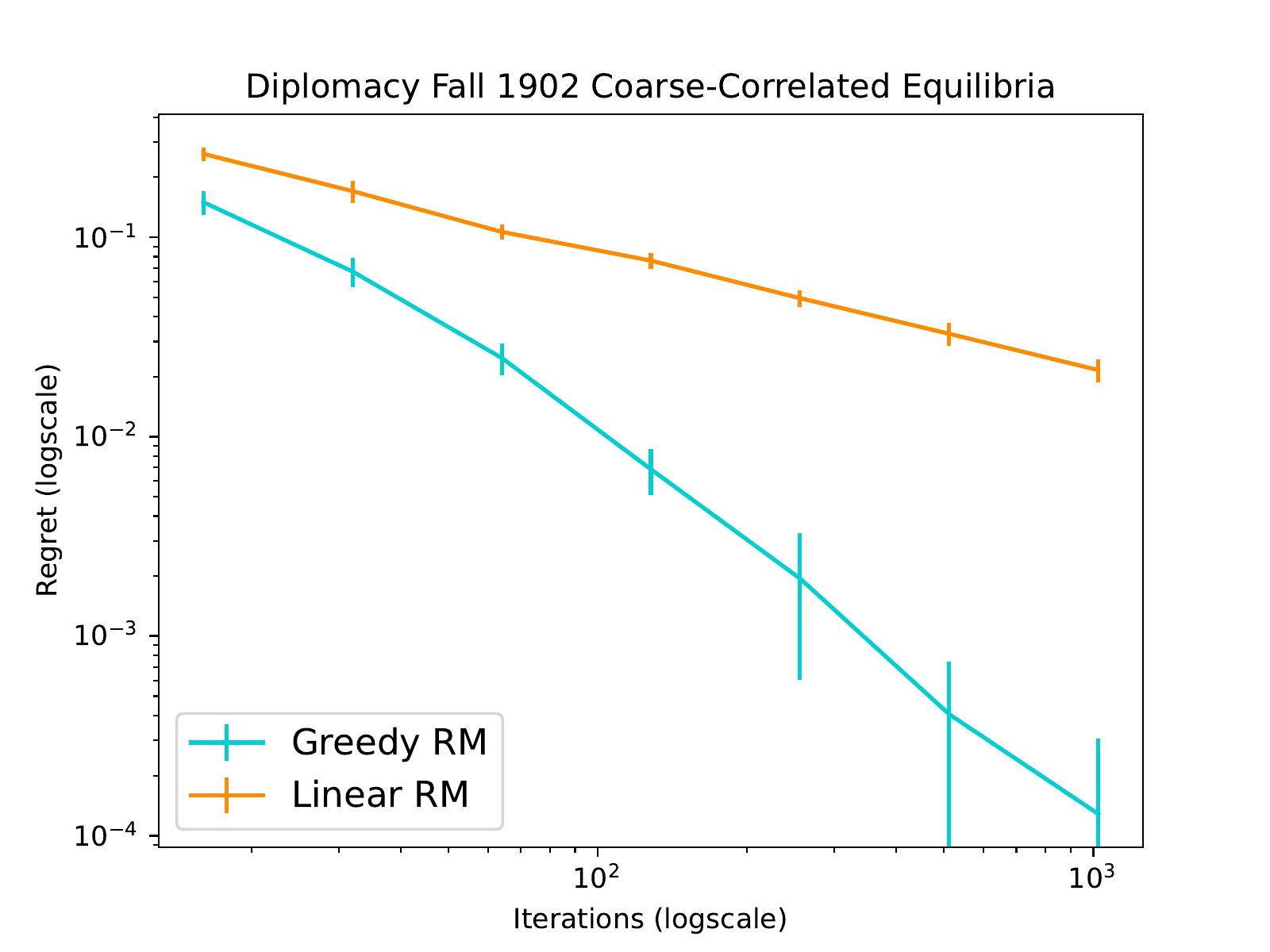}
  \caption{Midgame CCE Computation}
  \label{fig:fall_cce}
  \end{subfigure}
\centering
\begin{subfigure}[b]{.45\textwidth}
  \centering
  \includegraphics[width=\columnwidth]{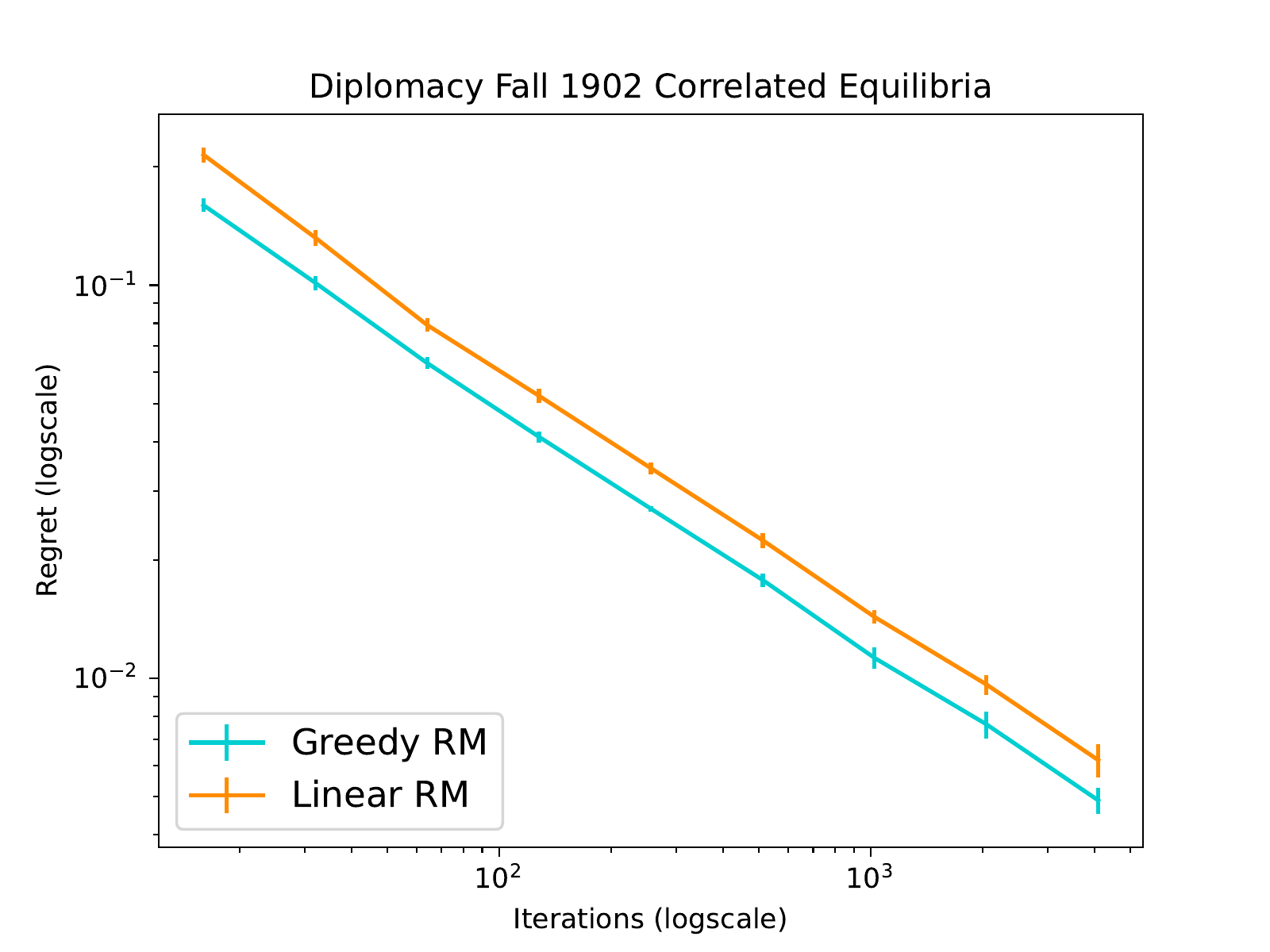}
  \caption{Midgame CE Computation}
  \label{fig:fall_ce}
  \end{subfigure}
\caption{We benchmark greedy weights for computing both CCE and CE on a turn in the middle of the game of Diplomacy, which are computed by minimizing external and internal regret respectively. Greedy weights is significantly faster than in all cases considered, and when computing CCE, we find gains of several orders of magnitude. Note that both axes are logscale. Error bars denote $95\%$ confidence intervals.
  }
\label{fig:diplomacyexperiments7Pappendix}
\vspace{-10pt}
\end{figure*}
\clearpage
\section{Regret Minimization with Greedy Weights on Normal-Form Games in the OpenSpiel Library}

\begin{figure}[!htb]
    \centering
    \includegraphics[width=0.45\columnwidth]{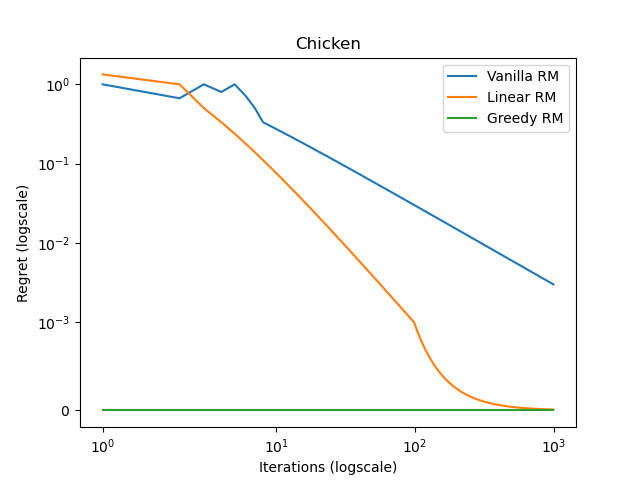}
    \includegraphics[width=0.45\columnwidth]{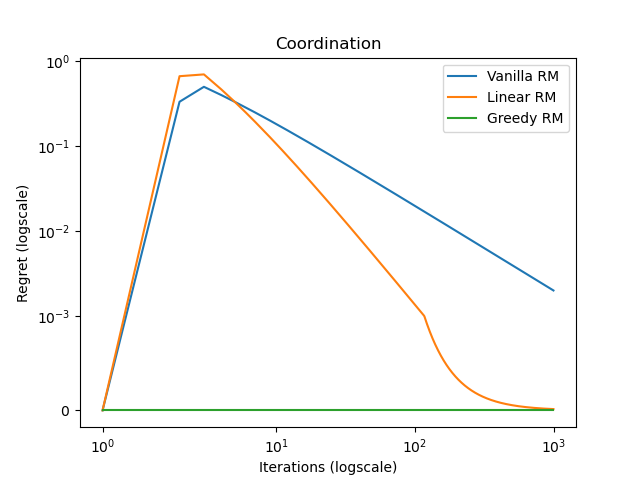}
    \includegraphics[width=0.45\columnwidth]{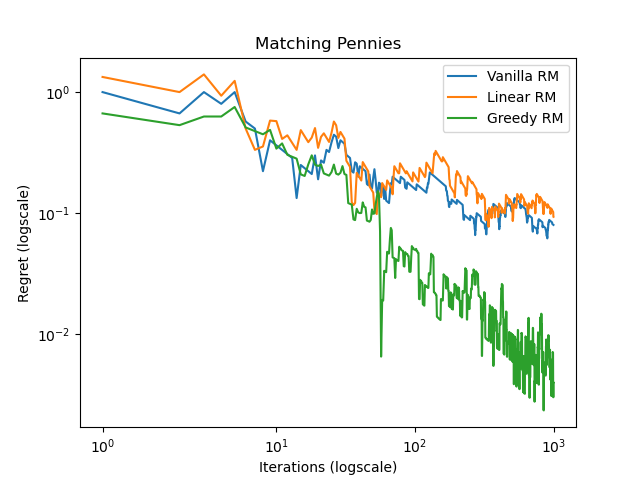}\ContinuedFloat
    \includegraphics[width=0.45\columnwidth]{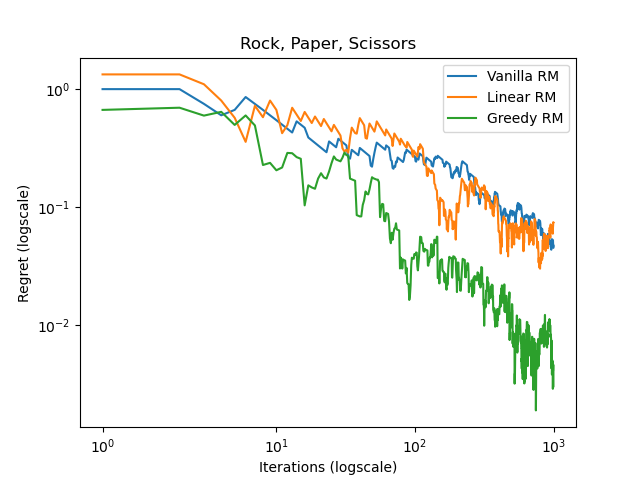}
    \includegraphics[width=0.45\columnwidth]{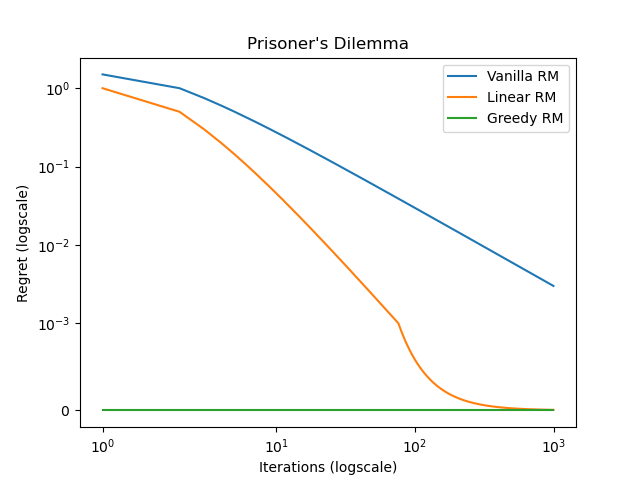}
    \includegraphics[width=0.45\columnwidth]{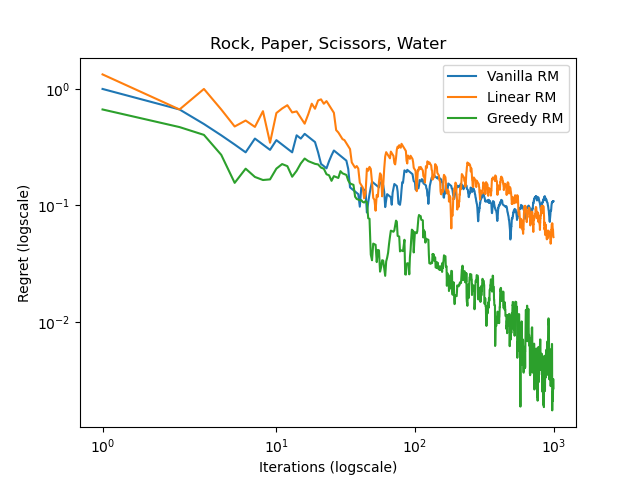}
\end{figure}
\begin{figure}[!htb]
    \centering
    \ContinuedFloat
    \includegraphics[width=0.45\columnwidth]{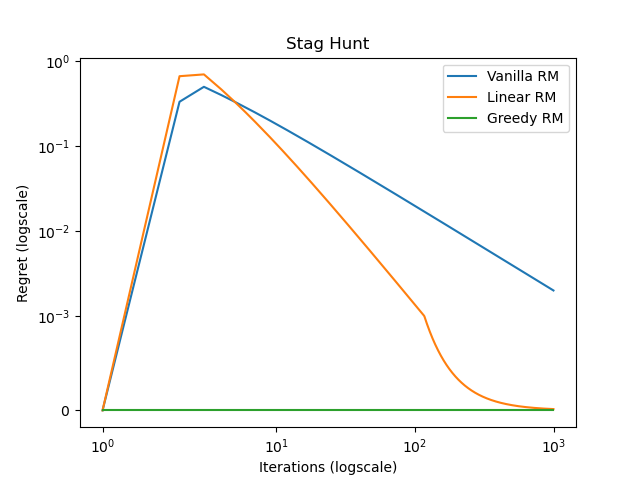}\ContinuedFloat
    \includegraphics[width=0.45\columnwidth]{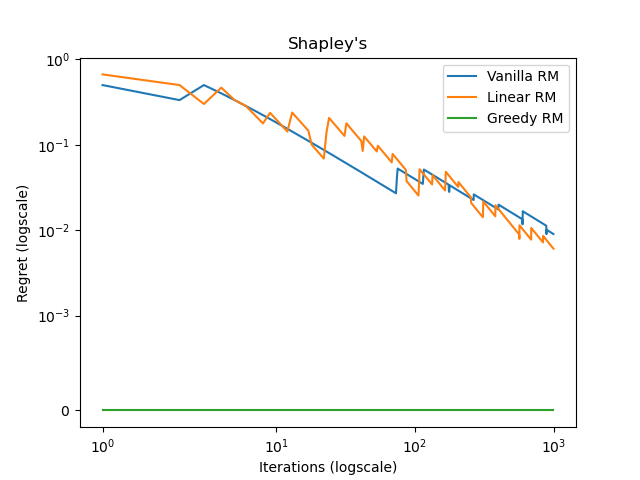}
     \includegraphics[width=0.45\columnwidth]{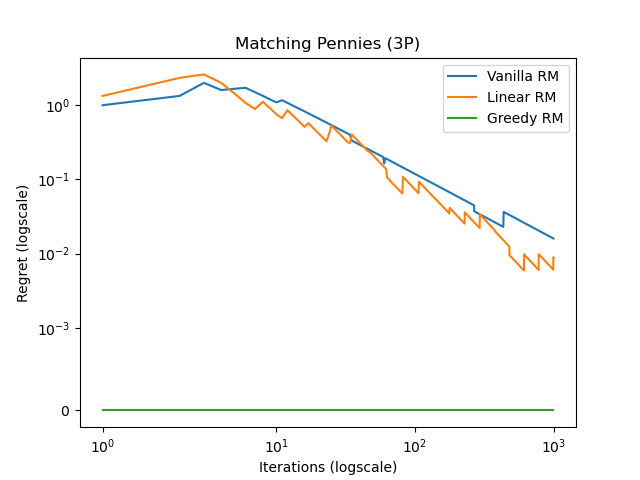}
    \caption{We run greedy weights on all real-world normal-form games included in the popular game theory library OpenSpiel \cite{lanctot2019openspiel} and find that greedy weights exceeds the performance of prior methods on all games evaluated.}
    \label{fig:openspielgames}
\end{figure}
\clearpage

\section{Description of Diplomacy}
In this section we summarize the rules of Diplomacy. A more detailed description can be found in~\cite{paquette2019no}.
No-press Diplomacy is a 7-player zero-sum board game in which players compete to control a majority of the 34 \emph{supply centers} (SCs). The map contains 75 locations, only some of which are SCs. At the start of the game, players control only 3-4 units and 3-4 SCs. At any point prior to a player gaining control of a majority of SCs, all remaining players may agree to a draw. In that case, the game ends and players receive a score of $C_i^2 /\sum_{i'}C_{i'}^2$, where $C_i$ is the number of SCs that player~$i$ controls.

Each turn, all players simultaneously assign one order to each of their units. Valid orders for a unit include holding the unit stationary, moving the unit to an adjacent location, supporting another unit's move, moving across water via a convoy, or convoying another unit across water.

In this paper we consider specifically the \emph{no-press} variant of Diplomacy, in which player cannot communicate. In the full game, all players may communicate via private pairwise cheaptalk channels. Since the game state in no-press Diplomacy is fully observable after each turn, the game can be viewed as a series of normal-form games, where the rows in the matrix corresponds to players' actions. Therefore, given a value function, each normal-form game can be viewed as its own subgame.

\clearpage

\newpage
\onecolumn
\section{Convergence Proof for Theorem~\ref{mainproposition}}
\label{sec:mainpropproof}


To simplify the proofs, we rescale the weights after each iteration so that $\sum_{t\leq T} w^t = T$. This has no effect on the strategies played or the weighted average regret. Let $A_I = \max_{i \in \mathcal{P}} |A_i|$.
\begin{proof}
Let $\Phi$ be the cumulative potential function for greedy weights on iteration $t$ for player~$i$. Let ${\bf R}_i^T$ be the regret vector for player~$i$ on iteration~$T$ and let $r_i^{T+1}$ be the new regret vector from iteration $T+1$ to be added to ${\bf R}_i^T$. In external regret matching, $$\Phi({\bf R}_i^T) = \sum_{a_i \in A_i} R_+^t(a_i)^2$$
In internal regret matching, $$\Phi({\bf R}_i^T) = \sum_{a_i^A, a_i^B \in A_i} R_+^T(a_i^A, a_i^B)^2$$
We will prove by induction that
\begin{equation}
\sum_{i \in \mathcal{P}} \Phi({\bf R}_i^T) \le |P| C T
\label{eqn:cond_regret}
\end{equation}
for a constant $C$ that depends only on the game and on the regret minimization algorithm used.

From \pyear{cesa2006prediction} (Theorem 2.1) we know that $\Phi({\bf R}_i^{T+1}) \le \Phi({\bf R}_i^{T}) + C$ whenever external or internal regret matching is played on iteration $T+1$. For external regret matching, $C = \Delta^2 |A_I|$. For internal regret matching, $C = \Delta^2 |A_I|^2$. This result holds regardless of the sequence of actions that took place up to $T$.

\begin{align}
 \sum_{i \in P} \Phi({\bf R}^T_i + r^{T+1}_i) & \leq |P|CT + |P|C \leq |P|C(T+1) \hspace{1cm} \\
\sum_{i \in P} \frac{\Phi({\bf R}^T_i + r^{T+1}_i)}{(T+1)^2} & \leq \frac{|P|C}{T+1} \\
\min_{w^{T+1}} \sum_{i \in P} \frac{\Phi({\bf R}^T_i + w^{T+1} r^{T+1}_i)}{(T + w^{T+1})^2} & \leq \frac{|P|C}{T + 1} \hspace{1cm} \textrm{($w^{T+1}$ selected by Algorithm~\ref{mainalgorithm})} \\
\sum_{i \in P} \frac{w'^2\Phi({\bf R}^T_i + w^{T+1} r^{T+1}_i)}{w'^2(T + w^{T+1})^2} & \leq \frac{|P|C}{T + 1} \hspace{1cm} \textrm{Rescaling }w' = \frac{T+1}{T+w^{T+1}} \label{eqn:reweight_generic} \\
\sum_{i \in P} \frac{\Phi(w' {\bf R}^T_i + w' w^{T+1} r^{T+1}_i)}{(T+1)^2} & \leq \frac{|P|C}{T + 1} \\
\sum_{i \in P} \Phi(w' {\bf R}^T_i + w' w^{T+1} r^{T+1}_i) & \leq |P|C (T + 1) \\
\label{eqn:weighted_generic}
\end{align}
\end{proof}

which satisfies Eq. (\ref{eqn:cond_regret}) for $T+1$. It's also clear that $\sum_{t\leq T+1} w^t = T+1$ from the reweighting procedure in (\ref{eqn:reweight_generic}). This completes the proof.

Finally, we can derive the weighted average regret. For external regret, this is:
\begin{align}
    \sqrt{ \sum_{a_i \in A_i} \max(0, \sum_{t\leq T} w^t r^t(a_i))^2} & \leq \sqrt{\sum_{i\in P} \sum_{a_i \in A_i} \max(0, \sum_{t\leq T} w^t r^t(a_i))^2} \\
    & \leq \sqrt{\Delta^2 |P| A_I T} \hspace{1cm} \textrm{(From (\ref{eqn:cond_regret}))}\\
    \sum_{a_i \in A_i} \max(0, \sum_{t\leq T} w^t r^t(a_i)) & \leq \sqrt{\Delta^2 |P| A_I T} \hspace{1cm} \textrm{(Jensen's inequality)}\\
    \sum_{a_i \in A_I} \sum_{t \leq T} w^t r^t(a_i) & \leq \sqrt{\Delta^2 |P| A_I T} \\
    \bar{R}_i^T(a_i) = \frac{\sum_{a_i \in A_i} \sum_{t\leq T} w^t r^t(a_i)}{\sum_{t\leq T} w^t} & \leq \sqrt{\frac{\Delta^2 |P| A_I}{T}}
\end{align}

For internal regret, this is:
\begin{align}
    \sqrt{ \sum_{a_i^A, a_i^B \in A_i} \max(0, \sum_{t\leq T} w^t r^t(a_i^A, a_i^B))^2} & \leq \sqrt{\sum_{i\in P} \sum_{a_i^A, a_i^B \in A_i} \max(0, \sum_{t\leq T} w^t r^t(a_i^A, a_i^B))^2} \\
    & \leq \sqrt{\Delta^2 |P| A_I^2 T} \hspace{1cm} \textrm{(From (\ref{eqn:cond_regret}))}\\
    \sum_{a_i^A, a_i^B \in A_i} \max(0, \sum_{t\leq T} w^t r^t(a_i^A, a_i^B)) & \leq \sqrt{\Delta^2 |P| A_I^2 T} \hspace{1cm} \textrm{(Jensen's inequality)}\\
    \sum_{a_i^A, a_i^B \in A_I} \sum_{t \leq T} w^t r^t(a_i^A, a_i^B) & \leq \sqrt{\Delta^2 |P| A_I^2 T} \\
    \bar{R}_i^T(a_i^A, a_i^B) = \frac{\sum_{a_i^A, a_i^B \in A_i} \sum_{t\leq T} w^t r^t(a_i^A, a_i^B)}{\sum_{t\leq T} w^t} & \leq \sqrt{\frac{\Delta^2 |P| A_I^2}{T}}
\end{align}

If each player's weighted average regret is bounded by $\epsilon$, then the weighted average policy is an $\epsilon$-equilibrium \cite{hart2000simple}.
\clearpage

\end{document}